\documentclass[a4paper,runningheads]{llncs}

\usepackage[utf8]{inputenc}
\usepackage[english]{babel}
\usepackage{microtype}
\usepackage{pifont}
\usepackage[autostyle]{csquotes}

\usepackage[inline]{enumitem}

\usepackage{hyperref}
\usepackage{bookmark}

\usepackage{enumitem}

\usepackage{amsmath}
\usepackage{amssymb}
\usepackage{mathtools}

\usepackage{booktabs}
\usepackage{makecell}
\usepackage{multirow}

\usepackage{subcaption}
\DeclareCaptionSubType*{figure}

\usepackage{tikz}
\usetikzlibrary{calc,fit,matrix,positioning,arrows,arrows.meta,decorations.markings}

\usepackage{bussproofs}
\usepackage{stackengine}
\EnableBpAbbreviations{}

\usepackage{algorithm2e}

\SetAlgoCaptionSeparator{.}
\SetAlCapSkip{8pt}

\newenvironment{myalgorithm}[1][htb]
{%
  \begin{algorithm}[#1]%
  }{\end{algorithm}}

\usepackage{tabularx}

\title{Relational Equivalence Proofs Between Imperative and MapReduce
  Algorithms}

\author{Bernhard Beckert \and Timo Bingmann \and Moritz Kiefer \and Peter Sanders \and Mattias~Ulbrich \and Alexander~Weigl}
\authorrunning{Beckert, Bingmann, Kiefer, Sanders, Ulbrich, Weigl}
\institute{
  Institute of Theoretical Informatics\\
  Karlsruhe Institute of Technology, Germany
}

\usepackage[useregional]{datetime2}
\date{\today}

\titlerunning{Equivalence of Imperative and MapReduce Algorithms}

\newcommand{\cxx}{C\nolinebreak\hspace{-.05em}\raisebox{.4ex}{\tiny\bf +}\nolinebreak\hspace{-.10em}\raisebox{.4ex}{\tiny\bf +}}
\newcommand{\java}{Java}
\newcommand{\steprel}{\Rightarrow_{\mathit{bs}}}
\newcommand{\steps}[2]{\ensuremath{\mathit{#1} \steprel{} \mathit{#2}}}
\newcommand{\equivalent}[3]{#2 \cong_{#1} #3}
\newcommand{\hastype}[3]{\ensuremath{#1 \vdash{} #2 : #3}}
\newcommand{\subst}[3]{[#1/#2]#3}

\newcommand{\TInt}{\mathrm{Int}}
\newcommand{\TBool}{\mathrm{Bool}}
\newcommand{\TList}[1]{[#1]}
\newcommand{\TProd}[2]{#1\times{} #2}
\newcommand{\TArrow}[2]{#1 \to{} #2}
\newcommand{\TSum}[2]{#1 + #2}
\newcommand{\TUnit}{\mathrm{Unit}}

\newcommand{\alt}{~~|~~}

\newcommand{\tapp}[2]{\mathsf{app} (#1, #2)}
\newcommand{\tabs}[2]{\lambda #1.\, #2}
\newcommand{\tpair}[2]{\mathsf{pair} (#1, #2)}
\newcommand{\tfst}[1]{\mathsf{fst} (#1)}
\newcommand{\tsnd}[1]{\mathsf{snd} (#1)}
\newcommand{\tinl}[1]{\mathsf{inl} (#1)}
\newcommand{\tinr}[1]{\mathsf{inr} (#1)}
\newcommand{\tcase}[3]{\mathsf{case}\: #1\: \mathsf{of}\: \tinl{l} \mapsto #2;\: \tinr{r} \mapsto #3}
\newcommand{\tmap}[2]{\mathsf{map} (#1,#2)}
\newcommand{\tzip}[2]{\mathsf{zip} (#1,#2)}
\newcommand{\titer}[2]{\mathsf{iter} (#1,#2)}
\newcommand{\tif}[3]{\mathsf{if}\: #1\: \mathsf{then}\: #2\: \mathsf{else}\: #3}
\newcommand{\tfold}[3]{\mathsf{fold} (#1,#2,#3)}
\newcommand{\tint}[1]{\mathsf{int} (#1)}
\newcommand{\tbool}[1]{\mathsf{bool} (#1)}
\newcommand{\tunit}{\mathsf{unit}}
\newcommand{\tlist}[1]{\mathsf{list}[#1]}
\newcommand{\treplicate}[2]{\mathsf{replicate} (#1,#2)}
\newcommand{\trange}[2]{\mathsf{range} (#1,#2)}
\newcommand{\tlength}[1]{\mathsf{length} (#1)}
\newcommand{\tconcat}[1]{\mathsf{concat} (#1)}
\newcommand{\tgroup}[1]{\mathsf{group} (#1)}
\newcommand{\tread}[2]{\mathsf{read} (#1,#2)}
\newcommand{\twrite}[3]{\mathsf{write} (#1,#2,#3)}
\newcommand{\treadatkey}[2]{\mathsf{readAtKey} (#1,#2)}
\newcommand{\twriteatkey}[3]{\mathsf{writeAtKey} (#1,#2,#3)}
\newcommand{\tadd}[2]{\mathsf{add}(#1,#2)}
\newcommand{\tsub}[2]{\mathsf{sub}(#1,#2)}
\newcommand{\tmul}[2]{\mathsf{mul}(#1,#2)}
\newcommand{\tgt}[2]{\mathsf{gt}(#1,#2)}
\newcommand{\tlt}[2]{\mathsf{lt}(#1,#2)}

\newcommand{\reffig}[1]{Fig.~\ref{fig:#1}}
\newcommand{\refsec}[1]{Sect.~\ref{sec:#1}}

\newcommand{\refalg}[1]{Alg.~\ref{alg:#1}}

\newcommand{\primop}[1]{\texttt{#1}}

\newcommand{\introparagraph}[1]{\paragraph{\upshape\bfseries #1.}}

\newcommand{\IL}{\textsf{IL}}
\newcommand{\FFL}{\textsf{FFL}}

\newcommand{\kmeans}{\emph{$k$-means}}
\newcommand{\pagerank}{\emph{PageRank}}

\newcommand{\sideconditions}[1]{\textbf{Side conditions:}\ #1}
\DeclareRobustCommand{\rewriterule}[3]{
  {
  \begin{tabular}[t]{l>{\centering\arraybackslash}p{1cm}l}
    #1 & \(\leftrightsquigarrow\) & #2
  \end{tabular}
  \def\temp{#3}\ifx\temp\empty
  \else
  {
  \\[.3em]
  \noindent
  \sideconditions{#3}
  }
  \fi
  }
}
\newcommand{\fv}[1]{\mathit{FV} (#1)}

\newcommand{\mapreduce}{\emph{MapReduce}}
\newcommand{\coq}{\emph{Coq}}

\newcommand{\numrules}{13}
\newcommand{\numprovenrules}{10}

\begin{document}

\mainmatter

\maketitle

\begin{abstract}
  \mapreduce{} frameworks are widely used for the implementation of
  distributed algorithms. However, translating imperative algorithms
  into these frameworks requires significant structural changes to the
  algorithm. As the costs of running faulty algorithms at scale can be
  severe, it is highly desirable to verify the correctness of the translation,
  i.e., to prove that the \mapreduce{} version is equivalent to the imperative
  original.

  We present a novel approach for proving equivalence between imperative and
  \mapreduce{} algorithms based on partitioning the equivalence proof into a
  sequence of equivalence proofs between intermediate programs with smaller
  differences.  Our approach is based on the insight that two kinds of
  sub-proofs are required: (1)~uniform transformations changing the
  control-flow structure that are mostly independent of the particular context
  in which they are applied; and (2)~context-dependent transformations that
  are not uniform but that preserve the overall structure and can be
  proved correct using coupling invariants.

  We demonstrate the feasibility of our approach by evaluating it on
  two prototypical algorithms commonly used as examples in \mapreduce{} frameworks: $k$-means
  and PageRank. To carry out the proofs, we use the interactive theorem
  prover \coq{} with partial proof automation. The results show that our
  approach and its prototypical implementation based on \coq{} enables
  equivalence proofs of non-trivial algorithms and could be automated to a
  large degree.
\end{abstract}

\setcounter{tocdepth}{3}

\section{Introduction}
\introparagraph{Motivation}
Frameworks such as \mapreduce{}~\cite{mapreduce}, Spark~\cite{spark}
and Thrill~\cite{Thrill} address the challenges arising in the
implementation of distributed algorithms by providing a limited set of
operations whose execution is automatically parallelized and
distributed among the nodes in a cluster. However, translating an
existing imperative algorithm into such a framework is a challenge in
itself and the original algorithmic structure is often lost during
that translation since imperative constructs do not translate directly
to the provided primitives. Implementing efficient
algorithms using \mapreduce{} frameworks can thus require significant
changes to the original algorithm.

By proving the equivalence of the original imperative algorithm and
its \mapreduce{} version, one can verify that no bugs have been
introduced during the translation. While such proofs do not directly
provide correctness guarantees for the \mapreduce{} algorithm, they
transfer correctness results from the imperative version to the
\mapreduce{} implementation.
The transferred correctness properties can be formal proofs whose
reach then extends to the distributed implementation, but can
also be informal arguments, e.g., if the algorithm is a well-known,
simple textbook reference implementation or if it has been
successfully applied previously.

In this paper, we use the term ``\mapreduce{}'' in a broader sense than
implied by the two functions ``map'' and ``reduce''. While some
frameworks such as Hadoop's MapReduce~\cite{white2012hadoop} module
are programmed strictly by specifying these two functions, the more
popular and widely used distributed frameworks provide many additional
primitives for performance reasons and to make them easier to program
with. Theoretically, these additional primitives can be reduced to
only map and reduce operations~\cite{chambers2010flumejava}, but this
overly complicates the program description and is generally not used
in real-world applications. In Appendix~\ref{sec:ffl-syntax}, we list
the few additional primitives we consider part of the extended
\mapreduce{} operation set.

\introparagraph{Contribution of this paper}

We present an interactive verification approach with which a \mapreduce{}
implementation of an algorithm can be proved equivalent to an imperative
implementation (to the best of our knowledge this is the first framework for
the purpose of such equivalence proofs, see \refsec{relatedwork}).
Proofs are conducted as chains/sequences of individual, smaller
behavior-preserving program transformations.

We came to the important insight that two kinds of sub-proofs are required:
(1)~uniform transformations changing the control-flow structure that are
mostly independent of the particular context in which they are applied; and
(2)~context-dependent transformations that are not uniform but that preserve
the overall structure and can be proved correct using coupling invariants.  We
identified and describe a catalogue of \numrules{} individual rules.
Correctness of \numprovenrules{} of those rules was proven formally using the
\coq{} theorem prover.

Our approach has a high potential for
automation. The required interaction is designed to be as high-level
as possible. The proof is guided by user-specified intermediate
programs from which the individual transformations are derived. The
rules are designed such that their side conditions can be proved
automatically and we describe how pattern matching can be used to
allow for a more flexible specification of intermediate steps.

We describe a workflow for integrating this approach with existing interactive
theorem provers. We have successfully implemented the approach as a prototype
within the interactive theorem prover \coq{}~\cite{Coq} and evaluated the
feasibility of our approach by applying it to the $k$-means and PageRank algorithms.
These two are prototypical algorithms commonly used as examples in
\mapreduce{} frameworks, because they exhibit the most common patterns
found in large-scale distributed data processing applications.
By showing that our approach can be applied to these two examples, we
demonstrate that it can be extended to a much larger set of applications.

\begin{figure}[t]
  \centering
  \begin{tikzpicture}[
    y=-1cm,
    block/.style={
      draw,
      minimum height=1cm,
      anchor=center,
      align=center,
    },
    textblock/.style={
      block,
      text width=1.45cm,
      font=\sffamily\scriptsize
    },
    decoration={markings,mark=at position 1 with
      {\arrow{To[length=.1cm]}}},
    font=\sffamily
    ]
    \matrix (highlevel-chain) [matrix of nodes,nodes in empty cells,column sep=.3cm]
    {
      |[textblock]| Imperative algorithm &
      |[draw=none,align=center]| \ldots &
      |[textblock]| A &[2cm]
      |[textblock]| B &
      |[draw=none,align=center]| \ldots &
      |[textblock]| MapReduce algorithm \\
    };
    \draw[<-] (highlevel-chain-1-1) to (highlevel-chain-1-2);
    \draw[->] (highlevel-chain-1-2) to (highlevel-chain-1-3);
    \draw[<-] (highlevel-chain-1-4) to (highlevel-chain-1-5);
    \draw[->] (highlevel-chain-1-5) to (highlevel-chain-1-6);
    \draw[<->] (highlevel-chain-1-3) to node[below] (highlevel-eq-1) {$\cong$} (highlevel-chain-1-4);
    \node[below=0cm of highlevel-chain-1-2.center] (highlevel-eq-0) {$\cong$};
    \node[below=0cm of highlevel-chain-1-5.center] (highlevel-eq-2) {$\cong$};

    \matrix (lowlevel-chain) [matrix of nodes,nodes in empty cells,column sep=.3cm,below=of highlevel-chain]
    {
      |[textblock]| Imperative algorithm &
      |[draw=none,align=center]| \ldots &
      |[textblock]| A &[2cm]
      |[textblock]| B &
      |[draw=none,align=center]| \ldots &
      |[textblock]| MapReduce algorithm \\
    };
    \draw[<-] (lowlevel-chain-1-1) to (lowlevel-chain-1-2);
    \draw[->] (lowlevel-chain-1-2) to (lowlevel-chain-1-3);
    \draw[<-] (lowlevel-chain-1-4) to (lowlevel-chain-1-5);
    \draw[->] (lowlevel-chain-1-5) to (lowlevel-chain-1-6);
    \draw[<->] (lowlevel-chain-1-3) to node[above] (lowlevel-eq-1) {$\cong$} (lowlevel-chain-1-4);
    \foreach \n in {1,3,4,6}{
      \draw[->] (highlevel-chain-1-\n) to (lowlevel-chain-1-\n);
    }
    \node[right,font=\sffamily\scriptsize,text width=width("translates")] at ($(highlevel-chain-1-3)!0.5!(lowlevel-chain-1-3)$) {translates to};
    \node[above=0cm of lowlevel-chain-1-2.center] (lowlevel-eq-0) {$\cong$};
    \node[above=0cm of lowlevel-chain-1-5.center] (lowlevel-eq-2) {$\cong$};

    \node[above=.3cm of highlevel-chain-1-6.north east,anchor=south east,align=right,inner sep=0] (highlevel) {Imperative Language (\IL)};
    \node[rounded corners,draw,fit=(highlevel-chain) (highlevel)] {};
    \node[below=.3cm of lowlevel-chain-1-6.south east,anchor=north east,align=right,inner sep=0] (lowlevel) {Formalized Functional Language  (\FFL)};
    \node[rounded corners,draw,fit=(lowlevel-chain) (lowlevel)] {};

    \newcommand{\doublearrow}[2]{
      \draw ([xshift=-.3mm]#1) -- ([xshift=-.3mm,yshift=-.49mm]#2);
      \draw ([xshift= .3mm]#1) -- ([xshift= .3mm,yshift=-.49mm]#2);
      \draw[draw=none,postaction={decorate}] (#1) -- (#2);
    }
    \foreach \n in {0,1,2}{
      \doublearrow{lowlevel-eq-\n.north}{highlevel-eq-\n.south};
    }
    \draw[draw=none] (lowlevel-eq-1) -- node[right,font=\sffamily\scriptsize] {implies} (highlevel-eq-1);
  \end{tikzpicture}
  \caption{Chain of equivalent programs is translated into formalized functional language}\label{fig:approach-overview}
\end{figure}
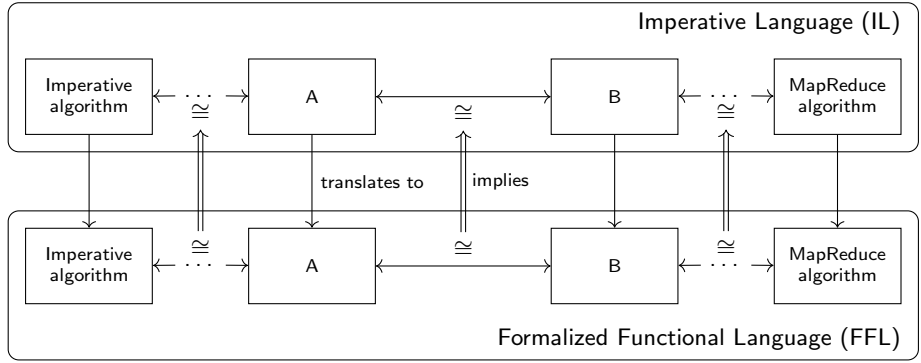

\introparagraph{Overview of the approach}

The main challenge in proving the equivalence of an imperative and a
\mapreduce{} algorithm lies in the potentially large structural
difference between two such algorithms.
Existing relational verification approaches
(like~\cite{rvt,automatingregver,symdiff,verdoolaege}) exploit the fact that
the two programs versions to be compared are structurally similar, which allows
the verification to focus on describing and proving the similarity of the
implementations rather than describing what they actually compute.
To deal with the complexity arising from the large structural
differences, the equivalence of imperative and \mapreduce{} algorithms
is not shown in one step, but as a succession of equivalence proofs
for structurally closer program versions.

To this end, we require that the translation of the algorithm is
broken down (by the user) into a chain of intermediate programs. For
each pair of neighboring programs in this chain, the difference is
comparatively small and can usually be reduced to one isolated
transformation.

The imperative algorithm, the intermediate programs, as well as
the \mapreduce{} implementation, are given in a
high-level imperative programming language~(\IL).
\IL{} is based on a while language and supports integers, booleans, fixed length arrays and sum and product types.
It does not support recursion.
Besides the imperative language constructs, \IL{} supports
\mapreduce{} primitives. Given that we have stated previously that
\mapreduce{} programs tend to be of a more functional nature, it might
seem odd at first to not use a functional language for specifying
\mapreduce{} algorithms. However, this is in accordance with the fact
that most of the existing \mapreduce{} frameworks are not implemented
as separate languages but as frameworks built on top of imperative
languages such as \java{}, Scala, or \cxx{}~\cite{spark,Thrill}. Thereby sequential parts of
\mapreduce{} algorithms can still be implemented using imperative
language features.

Each program specified in the high-level imperative language is then
automatically translated into the formalized functional language~(\FFL)
described in \refsec{formal-foundations}. The equivalence proofs are
conducted on programs in this functional language.
An overview of this process can be seen in \reffig{approach-overview}.
For each pair of neighboring programs in the chain, a proof obligation is
generated that requires proving their equivalence. These proof obligations are
then discharged independently of each other (using the workflow described in
\refsec{individual-equivalence-proofs}). Since, by construction, the semantics
of \IL{} programs is the same as that of corresponding \FFL{} programs, the
equivalence of two \IL{} programs follows from the equivalence of their
translations to \FFL{}.
Figure~\ref{fig:nonlocal-transformation} shows two example \IL{} programs for
calculating the element-wise sum of two arrays. %

\begin{figure}[t]
  \centering
  \SetKwProg{Fn}{Function}{\\begin}{end}
  \SetKwFunction{SumArrays}{SumArrays}
  \SetKwFunction{SumArraysZipped}{SumArraysZipped}
  \SetKwFunction{Zip}{zip}
  \SetKwFunction{Replicate}{replicate}
  \SetKwFunction{fst}{fst}
  \SetKwFunction{snd}{snd}
  \SetKwArray{zipped}{zipped}
  \SetKwArray{sum}{sum}
  \SetKwArray{xs}{xs}
  \SetKwArray{ys}{ys}
  \begin{minipage}[t]{.42\textwidth}
    \vspace{0pt}
    \begin{algorithm}[H]
      \Fn{\SumArrays{\xs,\ys}}{
        \sum \(\leftarrow\) \Replicate{n, 0}\;
        \For{$i\leftarrow 0$ \KwTo $n-1$}{
          $\sum{i} \leftarrow \xs{i} + \ys{i}$\;
        }
        \Return{\sum}\;
      }
    \end{algorithm}
  \end{minipage}
  \begin{minipage}[t]{.5\textwidth}
    \vspace{0pt}
    \begin{algorithm}[H]
      \Fn{\SumArraysZipped{\xs,\ys}}{
        \sum \(\leftarrow\) \Replicate{n, 0}\;
        \zipped $\leftarrow$ \Zip{\xs,\ys}\;
        \For{$i\leftarrow 0$ \KwTo $n-1$}{
          \sum{i} $\leftarrow$
          \fst{\zipped{i}} + \snd{\zipped{i}}\;
        }
        \Return{\sum}\;
      }
    \end{algorithm}
  \end{minipage}
  \caption{Two \IL{} programs which calculate the element-wise sum of two arrays.}\label{fig:nonlocal-transformation}
\end{figure}

The implementation of our approach based on the \coq{} theorem prover has only
limited proof automation and still requires a significant amount of
interactive proofs.
We are convinced, however, that our approach can be extended such that it
becomes highly automatised and only few user interactions or none at all are
required -- besides providing the intermediate programs.  Further challenges
include the extension of our approach to features such as references and
aliasing which are commonly found in imperative languages.

\introparagraph{Structure of this paper} In
Sec.~\ref{sec:formal-foundations}, we lay the formal groundwork for
our approach by defining the programming language used for equivalence
proofs and the notion of program equivalence used in this paper.
Sec.~\ref{sec:prog-trans} describes the two kinds of program
transformations that we have identified and the techniques for proving
equivalence using these transformations. The technical framework for
equivalence proofs and the potential for automation are described in
Sec.~\ref{sec:individual-equivalence-proofs} and their its
evaluation is in~Sec.~\ref{sec:experiments}. In
Sec.~\ref{sec:relatedwork}, we discuss work related to the ideas
presented in this paper. Finally, we conclude in
Sec.~\ref{sec:conclusion} and consider possible future work.

\section{Formal Foundations and Program Equivalence}
\label{sec:formal-foundations}

In this section, we briefly describe the language \FFL, introduce a reduction
big-step semantics for \FFL{} and discuss the notion of equivalence for
\FFL{} programs.
A full description of the syntax and semantics of \FFL{} can be found
in Appendix~\ref{sec:ffl-appendix}.

The primary design goal of \FFL{} is the capability to represent
both imperative and \mapreduce{} programs written in \IL.
To achieve this, we follow the work by
Radoi~et~al.~\cite{translatingimperative} and use a simply typed
lambda calculus extended by the theories of sums, products, and arrays.
Furthermore, the language also contains the programming primitives
usually found in \mapreduce{} frameworks.
We also want to limit the number of primitives included in \FFL{}
while still retaining expressiveness. This simplifies proving general
properties of \FFL{} and proving the correctness of rewrite rules. We
accomplish this by building upon the work of
Chen~et~al.~\cite{sparkspecification}, who describes how to reduce
the large number of primitives provided by \mapreduce{} frameworks to
a smaller core.

Two new primitives $\mathsf{iter}$ and
$\mathsf{fold}$ were added to translate imperative loops directly.
Compared to transforming imperative programs into a recursive form,
this allows a translation closer to the original program formulation.
The $\mathsf{fold}$ operator is used to translate bounded for-each iterator
loops into \FFL. The evaluation of the expression
$\mathsf{fold}\ f\ v_0\ \mathit{xs}$ starts with the initial loop
state $v_0$ and iterates over each value of the array $\mathit{xs}$
updating the loop state by applying $f$.
General while loops are translated using the $\mathsf{iter}$ function.
$\mathsf{iter}\ f\ v_0$ is evaluated by repeatedly applying $f$ to the loop
state (which is initially $v_0$) until $f$ returns $\tunit$ to
indicate termination.
Program terms incorporating $\mathsf{iter}$ need not evaluate to a
value since the construct allows formulating non-terminating programs.

The big-step operational reduction
semantics~\cite{bigstep} of \FFL{} is defined as a binary relation
$\steprel$. Note that, since \FFL{} is based on lambda
calculus, programs in \FFL{} as well as values are
\FFL\ expressions. The semantics predicate is thus a partial,
functional relation on \FFL{}-terms.
\begin{definition}%
  \label{def:evaluates-stuck}
  An \FFL\ term $t$ \emph{evaluates} to an \FFL\ term~$v$ if
  $\steps{t}{v}$ holds. A term~$t$ is called \emph{stuck} if there
  exists no~$v$ such that $\steps{t}{v}$. Terms that evaluate to
  themselves are called \emph{values}.
\end{definition}
A formal definition of the syntax of \FFL\ is given in form of typing rules in
Fig.~\ref{fig:langtyping}; and the semantics of \FFL{} is shown as
inference rules in Fig.~\ref{fig:langsemantics}, both in
Appendix~\ref{sec:ffl-appendix}.
The evaluation of a program $t$ in an input state (i.e., for an
argument tuple $a$) resulting in a output state $v$ (a result tuple)
can be formalized as the reduction evaluation of the application of
the program to the arguments:
$\langle t, a \rangle \Downarrow_{\mathit{bs}} v \;\;:=\;\;
\steps{\tapp t a}{v}$.

The semantics of \FFL{} is deterministic.  This may seem odd because most
\mapreduce{} frameworks take considerable leeway from fully deterministic execution in the
name of performance. For example, some operations may be evaluated in
a non-deterministic order depending on how fast data arrives over the network leading
to non-determinism if these operations are not commutative and associative.
However, non-determinism in \mapreduce{} algorithms is usually not
desired and the problem of checking whether or not a \mapreduce{} algorithm is
deterministic is orthogonal to proving that it is equivalent to an
imperative algorithm. We thus consider a deterministic language model
to be suitable for our purposes and defer checking of determinism to
other tools such as those developed by
Chen~et~al.~\cite{commutativityofreducers,commutativitytransducer}.

Since \FFL{} includes the potential for run-time errors such as
out-of-bound array accesses but does not include an explicit error
term, the step-relation $\steps{}{}$ is not total. The absence of an
explicit error term also has the consequence that one cannot
distinguish between non-termination and runtime errors according to the
definition of program equivalence in Definition~\ref{def:equivalence}.

The introduction of the semantics relation allows us to define a
notion of program equivalence for \FFL\ terms.

\begin{definition} \label{def:equivalence}
  Two well-typed \FFL\ terms $s$ and $t$ are called \emph{equivalent} if
  they (a)~are of the same type~$\tau$ and (b)~evaluate to the same
  values~$v$.
  We write $\equivalent \tau s t$ in this case.
  Using
  $\hastype{}{t}{\tau}$ to denote that the closed \FFL\ term $t$ has type
  $\tau$, this definition can be formalised as follows:
  \begin{equation}
    \label{eq:equivalence}
    \begin{array}{ll}
      \equivalent{\tau}{s}{t} ~~\coloneqq{}~~
      &\hastype{}{s}{\tau}\ \wedge
        \ \hastype{}{t}{\tau}\ \wedge {}\\
      &\forall \mathit{v}\ldotp (\steps{s}{v}) \Leftrightarrow (\steps{t}{v})
    \end{array}
  \end{equation}
\end{definition}

This definition of program equivalence also enforces \emph{mutual
  termination}~\cite{mutualtermination}, i.e., the property that
equivalent programs either both terminate or both diverge.
In particular, two non-terminating terms of the same type are
equivalent.

Note that the generated proof obligations require proving the
equivalence of functions applied to arbitrary inputs instead of
requiring proving the equivalence of functions themselves

\begin{figure}[t]
  \setlength{\belowcaptionskip}{-9pt}
  \centering
  \begin{tabular}{c@{\hskip 1cm}c}
    \(
    \arraycolsep=0pt
    \begin{array}[t]{ll}
      \mathsf{fold}( & \begin{array}[t]{ll}
                         \lambda & \mathit{sum}.\,\lambda i.\\
                                 & \twrite{\mathit{sum}}{i}{\mathit{xs}[i]+\mathit{ys}[i]},
                       \end{array} \\
                     & \treplicate{n}{0}, \\
                     & \trange{0}{n})
    \end{array}
    \)
    &
      \(
      \arraycolsep=0pt
    \begin{array}[t]{ll}
      \mathsf{snd}(\mathsf{iter}( & \begin{array}[t]{ll}
                                      \lambda & (i,\mathit{sum}). \\
                                              & \begin{array}[t]{p{.3cm}ll}
                                                  \(\mathsf{if}\) & i<n & \\
                                                                  & \mathsf{then}\: &\mathsf{inr}\, \begin{array}[t]{ll}
                                                                                                      ( & i+1,\\
                                                                                                        & \twrite{\mathit{sum}}{i}{\mathit{xs}[i]+\mathit{ys}[i]})
                                                                                                    \end{array} \\
                                                                  & \mathsf{else}\: &\mathsf{inl}\,\tunit{},
                                                \end{array}
                                    \end{array} \\
                                  & (0, \treplicate{n}{0})))
    \end{array}
    \)\\ \mbox{} \\
    \textbf{(a)} & \textbf{(b)}
  \end{tabular}

  \caption{Translation of function \texttt{SumArrays} (see
    \reffig{nonlocal-transformation}) into \FFL{} using \primop{fold} and
    \primop{iter} (where \(\mathsf{inl}\) and \(\mathsf{inr}\) denote the left
    and right injection into a sumtype).}\label{fig:sumarrays-ffl}
\end{figure}

\begin{example}
Figure~\ref{fig:sumarrays-ffl} shows two transformations of the
function \texttt{SumArrays} (see \reffig{nonlocal-transformation}) into
\FFL{}. In \reffig{sumarrays-ffl}~(a), the loop is translated using $\mathsf{fold}$, and in
\reffig{sumarrays-ffl}~(b) using the more general $\mathsf{iter}$.
In both cases, it can be observed that the local variables $i$ and
$sum$ become $\lambda$-bound variables of the translation
of the enclosing block, in this case the loop body.

The first translation has the initial state $\treplicate{n}{0}$, an
array of length $n$ with all values set to $0$, and it iterates over
the indices in the array ($[0;1;\ldots;n-1]$), updating the array
\emph{sum} in each iteration using the $\mathit{write}$ function of
the McCarthy theory of arrays.

The translation in \reffig{sumarrays-ffl}~(b) starts from the initial
loop state $(0, \treplicate{n}{0})$. In each iteration, an
\emph{if}-clause is used to check if the loop condition still
evaluates to \emph{true}. If that is the case, the index is
incremented and $\mathit{sum}$ is updated, otherwise the program exits
the loop as indicated by \(\mathsf{inl}\, \tunit\)
and evaluates to the current loop state.
\end{example}

\section{Program Transformations}\label{sec:prog-trans}

With the reduction of imperative and \mapreduce{} implementations to the
common language \FFL{}, we are able to prove
equivalence between two programs by constructing a chain of single,
isolated program transformations.
We categorize the transformations by their dependence on the
surrounding context.
A \emph{context-independent} transformation is an uniform
transformation as it replaces only one isolated subterm in the program
by an equivalent term.
This replacement has no effects on other parts of the program and has
only conditions on the replaced subterm.
In contrast, \emph{context-dependent} transformations do not replace
individual terms but require many small changes throughout different
parts of the programs.

For example,
consider the \IL{} programs in \reffig{nonlocal-transformation}.
In the left \IL{} program, the loop iterates over two separate arrays
$xs$ and $ys$ of the same length.
In the right \IL{} program, the loop iterates over a single array that
represents the \emph{zipped} version of $xs$ and~$ys$.
Inspection of the \FFL{} versions from \reffig{sumarrays-ffl} shows
that this program transformation requires two changes to individual
subterms: (a)~the initial loop state,
and (b)~adaption of the read and write references. %

We use two complementary techniques for proving the correctness of
a transformation depending on whether it is context-independent
or context-dependent: The equivalence of programs related by
context-independent transformations is proven using rewrite rules
(\refsec{rewrite-rules}) while the equivalence of programs related by
context-dependent transformation is shown using coupling predicates
(\refsec{coupling-predicates}).

\begin{figure}[t]
  \centering%
  \rewriterule%
  { \arraycolsep=0pt
    \(
    \begin{array}{ll}
      \mathsf{fold}( & \tabs{\mathit{acc}}{\tabs{x}{f(\mathit{acc},g(x))}}, \\
                     & i, \\
                     & \mathit{xs})
    \end{array}
    \)
  }
  { \arraycolsep=0pt
    \(
    \begin{array}{ll}
      \mathsf{fold}( & \tabs{\mathit{acc}}{\tabs{y}{f(\mathit{acc}, y)}}, \\
                     & i, \\
                     & \tmap{g}{\mathit{xs}})
    \end{array}
    \)
  }
  {\(\mathit{acc}\not\in\fv{f}, \mathit{x}\not\in\fv{f}, \mathit{y}\not\in\fv{f}, \mathit{x}\not\in\fv{g},\mathit{acc}\not\in\fv{g}\)}
  \caption{Rewrite rule for separating a loop body into two functions $f$ and $g$
    such that the evaluation of $g$ is independent of all other
    iterations and can be computed in parallel. $\fv g$ is the set of
    free, unbound variables in the term $g$.
  }\label{fig:foldmap}
\end{figure}

\subsection{Handling Context-Independent Transformations Using Rewrite
  Rules}\label{sec:rewrite-rules}

Intermediate programs are mostly linked by uniform
context-independent transformations on isolated subterms.
Instead of performing and proving these local transformations
manually, we can capture them into generalized rewrite rules.
That equivalence is preserved when these generalized rewrite rules
are applied, needs to be proven only once.
By maintaining and using a collection of local transformations that
have been proven correct, we can lower proof complexity and later
increase the computer assistance and automation.

A rewrite rule describes a bidirectional program transformation that
allows the replacement of a subterm within a program.
It is composed of two patterns and a set of side conditions which are
sufficient for the transformation to preserve program equivalence.
A pattern is an \FFL{} term containing metavariables.

To apply a rewrite rule on a program, we have to identify a subterm of
the program that (a)~matches the first pattern and (b) satisfies
the side conditions.
The transformed program is obtained by the instantiation of the other
pattern with the matched metavariables.
Since the sets of bound  metavariables in the two patterns can be
different, some metavariables may not be uniquely instantiated, leading
to a degree of freedom in the translation.
We will discuss the practical implications of this in
\refsec{proofs-rewriterules}.

While there is no hard limit on the complexity of the side
conditions that can be part of rewrite rules, it is desirable to use
side conditions that are simple and easy to check.
This prevents the application of rewrite rules from producing
auxiliary complex proofs due to complex side conditions.
In our experiments we only encountered the following three different kinds of
side conditions:
\begin{enumerate}
\item Two arrays $xs$ and $ys$ have the same length, i.e.,
  $\equivalent{\mathrm{int}}{\tlength{xs}}{\tlength{ys}}$.
\item $t$ is not stuck.
\item $x\not\in \fv{t}$ where $\fv{t}$ is the set of free variables in
  the term $t$.
\end{enumerate}
\refsec{proofs-rewriterules} discusses how these side conditions could
be discharged automatically.

To illustrate the kind of rewrite rules used in the equivalence proofs
described in this paper, we present two of the most commonly used
rewrite rules in detail. To demonstrate the feasibility of formal correctness
proofs for rewrite rules, we have
proven the correctness of most (\numprovenrules{} out of \numrules{} rules) of
our rules in \coq{}. A full listing of all rewrite rules can be found in Appendix~\ref{sec:rules-appendix}.

The first rule, shown in \reffig{foldmap}, decomposes the loop body of
a \primop{fold} expression into two separate functions $f$ and~$g$,
where $g$ is independent of other iterations.
Thus, $g$~can be computed in parallel using a \primop{map} operation.
This rewrite rule illustrates that rewrite rules used in proofs can
often also function as guidelines for parallelizing and distributing
imperative algorithms.

The second rule, shown in \reffig{groupsameindex}, is similar to the previous
rule in that it tries to separate independent parts of the loop body
so that they can be executed in parallel.
However, in this case, the part that is extracted is only independent
of other iterations that access different indices.
The \primop{group} operation can be used to group all accesses to the
same index.
Using \primop{map} one can then calculate the new values for each
index in $\mathit{xs}$ in parallel and update $\mathit{ys}$ with those
new values.

\begin{figure}[t]
  \centering
  \rewriterule%
  { \arraycolsep=0pt
    \(
    \begin{array}{ll}
      \mathsf{fold}( & \lambda\mathit{acc}.\,\lambda(i,x). \\
                     & \twrite{\mathit{acc}}{i}{f(i,x,\mathit{acc}[i])}, \\
                     & ys, \\
                     & \mathit{xs})
    \end{array}
    \)
  }
  { \arraycolsep=0pt
    \(
    \begin{array}{ll}
      \mathsf{fold}( &\lambda\mathit{acc}.\,\lambda(i,v).\, \twrite{\mathit{acc}}{i}{v}, \\
                     & \mathit{ys}, \\
                     & \begin{array}{ll}
                         \mathsf{map}( & \lambda(i,\mathit{vs}). \\
                                       & (i,\tfold{\tabs{x'}{\tabs{x}{f(i,x,x')}}}{\mathit{ys}[i]}{\mathit{vs}}), \\
                                       & \tgroup{xs}))
                       \end{array}
    \end{array}
    \)
  }
  {\(\mathit{acc}\not\in\fv{f},\mathit{x}\not\in\fv{f},\mathit{x'}\not\in\fv{f},\mathit{i}\not\in\fv{f},\mathit{vs}\not\in\fv{f}\)}
  \caption{Rewrite rule for grouping loop iterations which access the
    same index of an array.}\label{fig:groupsameindex}
\end{figure}

\subsection{Handling Context-Dependent Transformations Using Coupling
  Predicates}\label{sec:coupling-predicates}
While context-independent transformations are nicely handled using
rewrite rules, context-dependent transformations can usually not be
captured by patterns and simple side conditions.
Coupling predicates provide a flexible and effective solution to proving
the correctness of context-dependent transformations -- at the cost of
requiring more user interactions than rewrite rules.
The use of coupling predicates is based on the observation that
analyzing two loops in lockstep and proving that a relational
property, i.e., the coupling predicate, holds after each iteration is
sufficient to prove that it holds after the execution of both loops.
\reffig{fold-coupling-rule} shows the corresponding coupling invariant
rule for \primop{fold}.
For the purpose of presentation, we ignore the distinction between
syntactic terms and the values to which they evaluate. Besides this
rule for \primop{fold}, there is a similar rule for \primop{iter}.

\begin{figure}[tb]
  \begin{equation*}
    \begin{aligned}
      &              &  & \mathcal{C}(i_{0},i_{0}')     \\
      & \ \ \ \wedge &  &
      (\forall i, i', j.\:
      \mathcal{C}(\mathit{i}, \mathit{i'})
      \implies \mathcal{C}(f(i,xs[j]),f'(i',xs'[j]))) \\
      & \implies     &  & \mathcal{C}(\tfold{f}{i_{0}}{xs},\tfold{f'}{i_{0}'}{xs'})
    \end{aligned}
  \end{equation*}
  \caption{Coupling invariant rule for \primop{fold} for a coupling
    predicate $\mathcal{C}$. Free variables are implicitly
    universally quantified.}\label{fig:fold-coupling-rule}
\end{figure}

One compelling example for using coupling predicates
is
given in the beginning of this section. The presented program
transformation is provable equivalent with the coupling invariant rule from
\reffig{nonlocal-transformation}.
If these arrays are part of the accumulator in a \primop{fold} or
\primop{iter}, capturing this transformation by a rule patterns is not
possible: While the transformation of the initial accumulator value
can be captured using patterns, this is not sufficient since all
references to the accumulator in the loop body also need to be
updated.
These references can be nested arbitrarily deep inside the loop body
and there can be arbitrarily many references.
This makes it impossible to capture them by a single pattern which can
only bind a fixed number of variables and thereby only make a fixed
number of transformations.
To make matters worse, it is not even sufficient to just transform the
loop itself since the loops are not equivalent: the right loop evaluates to
two separate arrays while the other evaluates to an array of tuples.
It is thus necessary to prove the equivalence of the enclosing terms
under the assumption that the loop in one program evaluates to a tuple
of two arrays $\tpair{\mathit{xs}}{\mathit{ys}}$ while the other loop
evaluates to $\tzip{\mathit{xs}}{\mathit{ys}}$.
This assumption can then be proven correct using the coupling
predicate stating that this holds after each iteration.

Another commonly found transformation is the removal of unused
elements from a tuple representing the loop accumulator.
As it was the case for the previous transformation, the loops
themselves are not equivalent and it is necessary to prove enclosing
terms equivalent using the assumption that the values present in both
loop accumulators are equivalent.
As before, this assumption can be proven correct using a coupling
predicate which states that this holds after each iteration.

\section{Transformation Application Strategy}
\label{sec:individual-equivalence-proofs}

Splitting the translation into a chain of intermediate programs and
translating these into \FFL{} leaves us with the problem of proving
neighboring programs equivalent.
In order to reduce the amount of user interaction required to conduct
these basic equivalence proofs, we define an iterative heuristic
search strategy to identify the locations within the programs on which
the program transformations described in Sect.~\ref{sec:prog-trans}
will be applied.
\refalg{workflow} depicts this search strategy as pseudocode. First, we
use the structural difference operation (\texttt{Diff}, see
\refsec{congruence}) to identify subterms $P'$ and $Q'$ whose
equivalence implies the equivalence of the full programs $P$ and $Q$.
Second, we start an iterative bottom-up process in which we try to
prove the equivalence of the subterms $P'$, $Q'$ and their enclosing
terms (\texttt{ProveEquivalent}), until we reached the top level
programs $P$ and~$Q$.
During the bottom-up process, the subterms $P'$ and $Q'$ may be found to be
equivalent only in some cases but not in others. But that is fine as long as
we are able to prove that the cases in which they are non-equivalent are not
relevant in the context in which $P'$ and $Q'$ occur.
Thus, we extract the premises under which $P'$ and $Q'$ are equivalent, and
bubble them up to the equivalence proof for the parent terms
(\texttt{AddMissingPremises}, \texttt{Widen}, see
Sec.~\ref{sec:missing-assumptions})
If we arrive at the top-level terms and cannot prove those
  equivalent, the proof fails. %

\begin{myalgorithm}[t]
  \SetKwInOut{Input}{input}
  \SetKwInOut{Output}{output}
  \SetKwFunction{Diff}{Diff}
  \SetKwFunction{ProveEquivalent}{ProveEquivalent}
  \SetKwFunction{AddMissingPremises}{AddMissingPremises}
  \SetKwFunction{Widen}{Widen}
  \SetKwData{premises}{Premises}
  \Input{Two \FFL{} terms $P$ and $Q$}
  \Output{\emph{true} if $P$ and $Q$ could be proven equivalent}
  \premises $\leftarrow$ \{\}\;
  ($P'$,$Q'$) $\leftarrow$ \Diff{$P$,$Q$}\;
  \Repeat{$P'=P$ and $Q'=Q$}{
    equivalent? $\leftarrow$ \ProveEquivalent{$P'$,$Q'$,\premises}\;
    \eIf{equivalent?}{
      \Return{true}\;
    }{
      \premises $\leftarrow$ \AddMissingPremises{\premises}\;
      $(P',Q') \leftarrow$ \Widen{$P'$,$Q'$}\;
    }
  }
  \Return{false}\;%
  \caption{Strategy for individual equivalence proofs between a pair
    of \FFL{} programs.}
  \label{alg:workflow}
\end{myalgorithm}

\subsection{Using Congruence Rules to Simplify
  Proofs}\label{sec:congruence}

While the difference between neighboring programs in the chain -- which are more
closely related -- tends to be small, the size of these programs can still be
large. This complicates interactive proofs for the user, and can also
slow down automated proofs. To reduce the complexity, we prove the
equivalence of subterms and then use congruence rules to derive the
equivalence of the full programs. A concrete example of a congruence
rule is shown in \ref{fig:congruence-rule}.

We have found that a simple structural comparison (\texttt{Diff} in
\refalg{workflow}) is well suited for finding smaller subterms whose
equivalence implies the equivalence of the full programs.
\texttt{Diff} computes the smallest two subterms such that replacing
them by placeholders results in identical terms. An example of
\texttt{Diff} can be seen in \ref{fig:diff-example}.

\begin{figure}[t]
  \begin{subfigure}[b]{.5\textwidth}
    \centering
    \AxiomC{\(\equivalent{\TList{\alpha}}{xs}{ys}\)}
    \AxiomC{\(\equivalent{\TInt}{i}{j}\)}
    \BinaryInfC{\(\equivalent{\alpha}{\tread{xs}{i}}{\tread{ys}{i}}\)}
    \DisplayProof{}
    \caption{Congruence rule for \texttt{read}}\label{fig:congruence-rule}
  \end{subfigure}\begin{subfigure}[b]{.5\textwidth}
    \centering
    \( \arraycolsep=0pt
    \begin{array}{ll}
      \texttt{Diff}( & \tfold{\tabs{(x,y)}{x+y}}{0}{xs}, \\
                     & \tfold{\tabs{(x,y)}{y+x}}{0}{xs}) \\
      \multicolumn{2}{l}{= (\tabs{(x,y)}{x+y}, \tabs{(x,y)}{y+x})}
    \end{array}
    \)
    \caption{Example of applying \texttt{Diff}}\label{fig:diff-example}
  \end{subfigure}
\end{figure}

\subsection{Missing Premises and Widening}
\label{sec:missing-assumptions}
During the iterative bottom-up process in \refalg{workflow},
$P'$~and $Q'$ may turn out to be non-equivalent in some cases.
The strategy then tries to extract required contextual conditions (premises)
that
are sufficient to ensure equivalence of $P'$ and~$Q'$
(\texttt{AddMissingPremises}).
In the next step, we try to prove the equivalence of enclosing terms
(\texttt{Widen}), which contain $P'$ and~$Q'$ as subterms.
Additionally, in the widening-step,
we take care of the generated premises.
These have either to be shown to always hold in the context of
$\texttt{Widen}(P',Q')$ or in the context of further widening.

\begin{figure}[b]
  \centering
  \begin{minipage}[t]{.4\textwidth}
    \vspace{0pt}
    \begin{algorithm}[H]
      \SetKwFunction{zip}{zip}
      \SetKwFunction{fst}{fst}
      \SetKwFunction{F}{F'}
      \SetKwArray{zipped}{zipped}
      \SetKwArray{sum}{sum}
      \SetKwArray{xs}{xs}
      \SetKwArray{ys}{ys}
      \sum $\leftarrow$ 0\;
      \For{$i\leftarrow 0$ \KwTo $n-1$}{
        \sum $\leftarrow$ \sum $+$ \xs{i}\;
        \xs $\leftarrow$ \F{\xs,\ys}\;
        }
    \end{algorithm}
  \end{minipage}
  \begin{minipage}[t]{.5\textwidth}
    \vspace{0pt}
    \begin{algorithm}[H]
      \SetKwFunction{zip}{zip}
      \SetKwFunction{fst}{fst}
      \SetKwFunction{F}{F}
      \SetKwArray{zipped}{zipped}
      \SetKwArray{sum}{sum}
      \SetKwArray{xs}{xs}
      \SetKwArray{ys}{ys}
      \sum $\leftarrow$ 0\;
      \For{$i\leftarrow 0$ \KwTo $n-1$}{
        \zipped $\leftarrow$ \zip{\xs,\ys}\;
        \sum $\leftarrow$ \sum $+$ \fst{$\zipped{i}$}\;
        \xs $\leftarrow$ \F{\zipped}\;
      }
    \end{algorithm}
  \vspace*{-2ex}
  \end{minipage}
  \caption{Two potentially equivalent \IL{} programs operating on two
    separate arrays (left) and the result of applying \texttt{zip} to
    these arrays (right). \texttt{xs}, \texttt{ys} are arrays of
    length $n$. \texttt{F} and \texttt{F'} return arrays of the length
    of their input.}\label{fig:strengthen-same-length}
\end{figure}

These two steps -- premise extraction and widening -- are commonly
required to prove the equivalence of loop bodies.
The example in \reffig{strengthen-same-length} illustrates this.
Applying \texttt{Diff} instantiates $P'$ and~$Q'$ with the
two loop bodies,  as they are the topmost non-equal subterms.
A coupling invariant implying that the two loops are started in
equivalent is not sufficient to ensure
equivalent loop states after execution
since \texttt{zip} is only defined for arrays of the same length.
Thus, the coupling invariant needs to include the premise that
 \texttt{xs} and \texttt{ys} are of equal length.

In some cases, additional premises sufficient for proving
equivalence can be found by working backward from missing assumption
in failed proofs.
In the example above, proving that the program states are equivalent
at the end of each loop iteration assuming that they are equivalent at
the beginning will fail due to the missing premise that \texttt{xs}
and \texttt{ys} have the same length. We thus add this premise and try
to prove the loop bodies equivalent using that premise. If that is
successful, we widen the context to enclosing terms. In the outer
context, we attempt to prove that the additional premises are
satisfied and derive the equivalence of the full loops based on proved
coupling invariant.

\subsection{Potential for Automation of Proofs using Rewrite Rules}\label{sec:proofs-rewriterules}
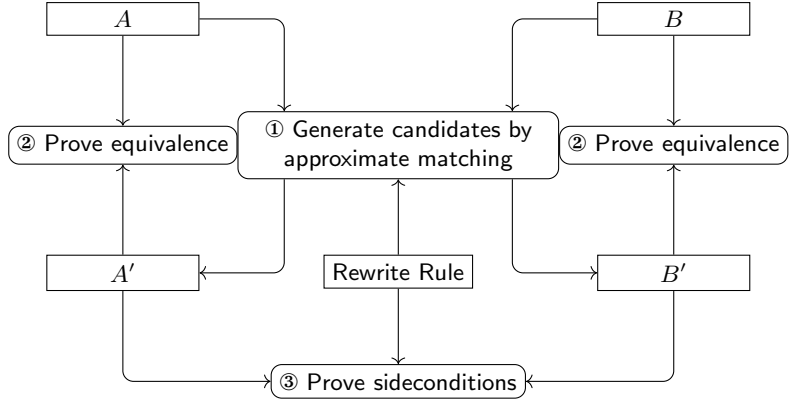
\begin{figure}[t]
  \begin{center}
    \begin{tikzpicture}[
      y=-1cm,
      edge/.style={->,rounded corners},
      input/.style={draw,align=center},
      process/.style={input,rounded corners},
      font=\sffamily
      ]

      \node[process,text width=4cm] (match) {
        \ding{172} Generate candidates by\\ approximate matching};
      \node[input,above left=1cm and .5cm of match,minimum width=2cm] (a) {$A$};
      \node[input,above right=1cm and .5cm of match,minimum width=2cm] (b) {$B$};
      \node[input,below=1cm of match] (rewriterule) {Rewrite Rule};
      \node[input,below left=1cm and .5cm of match,minimum width=2cm] (a') {$A'$};
      \node[input,below right=1cm and .5cm of match,minimum width=2cm] (b') {$B'$};
      \draw[edge] (rewriterule) -- (match);
      \draw[edge] (a) -- ([xshift=-1.5cm]a-|match) -- ([xshift=-1.5cm]match.north);
      \draw[edge] (b) -- ([xshift=1.5cm]b-|match) -- ([xshift=1.5cm]match.north);
      \draw[edge] ([xshift=-1.5cm]match.south) -- ([xshift=-1.5cm]a'-|match.south) -- (a');
      \draw[edge] ([xshift=1.5cm]match.south) -- ([xshift=1.5cm]b'-|match.south) -- (b');
      \node[process] at (a|-match) (eq-a) {\ding{173} \textsf{Prove equivalence}};
      \node[process] at (b|-match) (eq-b) {\ding{173} \textsf{Prove equivalence}};
      \draw[edge] (a) -- (eq-a);
      \draw[edge] (a') -- (eq-a);
      \draw[edge] (b) -- (eq-b);
      \draw[edge] (b') -- (eq-b);
      \node[process,below=1cm of rewriterule] (sideconditions)
      {\ding{174} Prove sideconditions};
      \draw[edge] (rewriterule) -- (sideconditions);
      \draw[edge] (a') -- (a'|-sideconditions) -- (sideconditions);
      \draw[edge] (b') -- (b'|-sideconditions) -- (sideconditions);
    \end{tikzpicture}
  \end{center}
  \caption{Workflow for equivalence proofs using rewrite rules. The
    user has to provide the programs $A,A',B,B'$ and also the
    rewrite rule. The equivalence proofs \ding{174} are
    computer-aided in \coq{}.}\label{fig:rewriterule}
\end{figure}

Since equivalence proofs using rewrite rules are particularly common
but also quite repetitive, this section is devoted to their potential for proof automation.
A graphical overview of the individual steps can be found in
\reffig{rewriterule}.
\begin{enumerate}
\item We perform an approximate matching procedure to generate
  candidate programs which match the patterns in the rewrite rule.
\item We attempt to prove that these candidates are equivalent
  to the input programs or otherwise we reject them.
\item We prove that the side conditions hold for these
  candidates.
\end{enumerate}
By the correctness of the rewrite rule, the candidates are equivalent.

\subsubsection{Matching of Rewrite Rules}
While automatic rewriting systems have been used in the related
context of automatically translating imperative algorithms to
\mapreduce{} algorithms~\cite{translatingimperative}, the specific ways
in which rewrite rules are used in our approach brings new challenges
as well as simplifications.

The challenge lies in the fact that the intermediate programs often do
not match the patterns found in rewrite rules directly.
There are two typical solutions: normal forms and generalization of
patterns.
Both are not applicable here.
First, there is no suitable normal form of \FFL{} programs.
Additionally, both programs are defined by the user, so we cannot
assume a specific program structure.
Second, the formulation of generalized rewrite rules for matching the
large variety of user-defined programs is difficult to obtain and also
their correctness proofs are harder to obtain.

The benefit of the programs $A,A'$ (resp.~$B,B'$) being
provided by the user is that this can reduce ambiguities.
In particular, the schematic variables in the two patterns usually
overlap to a large degree, but not fully.
The matching of the program $A$ against the corresponding pattern can
lead to unassigned metavariables, which we need to instantiate
with correct choices to prove the equivalence.
Now, we have the benefit, that the target program $A'$ is also defined
by the user.
So, we can obtain missing assignments by matching the other pattern
against the other program.

To find the intermediate programs which match the patterns in rewrite
rules, an approximate match procedure is used to find assignments for
schematic variables in patterns.
The approximate matching is an extension of the classical pattern with
the background knowledge and heuristic of easy-to-prove differences.
Applying these assignments to patterns yields candidates for the
intermediate program. Once two candidates that match the patterns in
a rewrite rule have been identified, it is necessary to prove that
\begin{enumerate*}[label=(\alph*)]
\item the candidates are equivalent to the programs used as the input
  of the approximate matching procedure, and
\item the side conditions hold and the equivalence of the candidates
  follows thereby from the correctness of the rewrite rule.
\end{enumerate*}.

While we have only implemented rudimentary partial automation of the
equivalence proof construction,
analyzing the \coq{} proofs produced in our experiments has shown that
these proofs can be reduced to the correctness of a small number of
simple transformations. Proving the correctness of these
transformations automatically is feasible and could drastically reduce
the need for user interaction.

During our evaluation, one of the most prevalent transformations is
\emph{call-by-name beta-reduction} or \emph{lambda abstraction}
depending on the direction of the transformation for proving the
equivalence (\ding{173}~in Fig.~\ref{fig:rewriterule}).
Call-by-name beta-reduction refers to the \emph{beta-reduction} found
in programming languages with lazy semantics, which contrary to
\emph{call-by-value beta-reduction} does not evaluate the argument
before applying substitution. Since we are working in a call-by-value
setting, call-by-name beta-reduction does not always produce an
equivalent program. However, the resulting program is equivalent if,
for each case where the argument would have been evaluated in the
original program, all occurrences in the new program will also be
evaluated.

Most other transformations are special cases of constant-folding,
e.g., reducing expressions such as \(\tfst{\tpair{a}{b}}\) to \(a\).
Constant-folding does not produce an equivalent program in general if
the terms that are being folded are inside the body of a lambda.
A sufficient criterion for the resulting program to be equivalent is
that the terms being folded are always evaluated.

\subsubsection{Proving Side Conditions}
In \refsec{rewrite-rules} we listed the three different kinds of side
conditions used in our rewrite rules. The first of those,
\(x\not\in\fv{t}\), is purely syntactical and can easily be checked
automatically. While proving that a term is not stuck can be difficult in
general, in our experiments this could usually be reduced to the term
being a value, which again is a syntactical condition. The third kind of side
condition states that two arrays have the same length. This can
usually be proven recursively by reduction to operations that
produce arrays of a specific length, e.g.,
\[ \forall n, a, b.\,
\tlength{\treplicate{n}{a}}=\tlength{\treplicate{n}{b}} \enspace, \]
or to
length-preserving operations such as $\mathsf{map}$. Note that it can be
necessary to strengthen loop invariants to carry this fact through a
loop as explained in \refsec{missing-assumptions}.

\section{Evaluation and Case Study}\label{sec:experiments}
To demonstrate the feasibility of our approach, we have created
a toolchain. The user to specifies a sequence of intermediate
programs in a simple imperative language. These programs are then
automatically translated into a formalization of the previously
described functional programming language \FFL\ in \coq{}. In addition to
generating proof obligations, our toolchain reduces these
obligations using the mentioned structural comparison
\texttt{Diff}, and it applies congruence rules to reconstruct an
equivalence proof of the full programs.

Using this toolchain, we have proven the equivalence of imperative and
\mapreduce{}s implementations of the \pagerank{} algorithm~\cite{pagerank} and
the \kmeans{}~\cite{lloyd1982least} algorithm in \coq{}.
\reffig{pagerank} shows the imperative and the \mapreduce{}
implementation of \pagerank{} that we have used in our experiments.

While we have created the imperative implementations
ot the two algorithms ourselves, the
\mapreduce{} versions are very close to the implementations
accompanying the Thrill~\cite{Thrill} framework. This reinforces our
claim that \FFL{} is capable of representing \mapreduce{} algorithms
and is thereby suitable for this approach. In total,
the formalization of \FFL{},
the rewrite rules, and proofs of various properties,
encompasses about 8000~lines of \emph{\coq{}} code. The equivalence
proofs of \pagerank{} and \kmeans{} each require about 3700~lines of
\coq{} proofs. That includes the automatically generated translation
of the chain of equivalent programs (for \kmeans{} this chain consists
of 9~programs while for \pagerank{} it consists of 6 programs), which
take up large parts of these proofs. The proofs rely on the rewrite
rules which we have formalized in \coq{} as well as coupling
predicates.

\begin{figure}[t]
  \SetAlFnt{\scriptsize}
  \newcommand{\algkwfont}[1]{\textbf{\sffamily{#1}}}
  \SetKwSty{algkwfont}
  \SetDataSty{texttt}
  \SetKwProg{Fn}{Function}{\\begin}{end}
  \SetKwFunction{PageRank}{PageRank}
  \SetKwFunction{Replicate}{Replicate}
  \SetKwFunction{Length}{Length}
  \SetKwFunction{Dampen}{Dampen}
  \SetKwFunction{Zip}{Zip}
  \SetKwFunction{Reduce}{Reduce}
  \SetKwFunction{Map}{Map}
  \SetKwFunction{FlatMap}{FlatMap}
  \SetKwArray{links}{links}
  \SetKwArray{ranks}{ranks}
  \SetKwArray{newRanks}{ranks'}
  \SetKwArray{outRanks}{outRanks}
  \SetKwArray{contribs}{contribs}
  \SetKwArray{rankUpdates}{rankUpdates}
  \SetKwData{iterations}{n}
  \SetKwData{contribution}{contrib}
  \SetKwData{numLinks}{numLinks}
  \SetKwData{r}{r}
  \SetKwData{i}{i}
  \SetKwData{l}{l}
  \SetKwData{ls}{ls}
  \SetKwData{p}{p}
  \SetKwData{q}{q}
  \SetInd{3pt}{6pt}
  \begin{minipage}[t]{.49\textwidth}
    \vspace{0pt}
    \hspace{-.5cm}
    \begin{algorithm}[H]
      \Fn{\PageRank{\links, \numLinks, \iterations}}{
        \ranks \(\leftarrow\) \\
        \Indp \Replicate{\numLinks, \(\frac{1}{\numLinks}\)}\;
        \Indm
        \For{\(\i=1\) \KwTo \iterations}{
          \newRanks \(\leftarrow\) \Replicate{\numLinks, \(0\)}\;
          \For {\(\p=0\) \KwTo{} \(\numLinks-1\)}{
            \contribution \(\leftarrow\) \(\frac{\ranks{\p}}{\Length{\links{\p}}}\)\;
            \ForEach{\(\q \leftarrow\) \links{\p}}{
              \newRanks{\q} \(\leftarrow\) \\
              \Indp \newRanks{\q} + \contribution\;
              \Indm
            }
          }
          \For{\(\p=0\) \KwTo{} \(\numLinks-1\)}{
            \ranks{\p} \(\leftarrow\) \\
            \Indp \Dampen{\newRanks{\p}, \numLinks};
          }
        }
        \Return{\ranks}\;
      }
    \end{algorithm}
  \end{minipage}
  \begin{minipage}[t]{.48\textwidth}
    \vspace{0pt}
    \begin{algorithm}[H]
      \Fn{\PageRank{\links, \numLinks, \iterations}}{
        \ranks \(\leftarrow\) \\
        \Indp \Replicate{\numLinks, \(\frac{1}{\numLinks}\)}\;
        \Indm
        \For{\(\i=1\) \KwTo \iterations}{
          \outRanks \(\leftarrow\) \Zip{\links,\ranks}\;
          \contribs \(\leftarrow\) \\
          \Indp \texttt{FlatMap}( \\
          \Indp \(\lambda(\ls,\r).\) \\
          \Indp \Map{\(\lambda \l. (\l, \frac{\r}{\Length{\ls}})\), \ls}, \\
          \Indm
          \outRanks)\;
          \Indm
          \Indm
          \rankUpdates \(\leftarrow\) \\
          \Indp \Reduce{\(+\), \(0\), \contribs}\;
          \Indm
          \newRanks \(\leftarrow\) \Replicate{\numLinks, \(0\)}\;
          \ForEach{\((\l,\r) \leftarrow \rankUpdates\)}{
            \newRanks{\l} \(\leftarrow \r\)\;
          }
          \ranks \(\leftarrow\) \\
          \Indp \Map{\(\lambda \r.\) \Dampen{\r, \numLinks}, \\
            \ \ \ \ \ \newRanks}\;
          \Indm
        }
        \Return{\ranks}\;
      }
    \end{algorithm}
  \end{minipage}
  \caption{Imperative (left) and \mapreduce{} (right) versions of the
    \emph{PageRank} algorithm (the function
    \texttt{Replicate(n,v)} creates an array
    of length~\texttt{n} with all elements set to~\texttt{v}; and \texttt{Dampen} is an arbitrary function).}\label{fig:pagerank}
\end{figure}

\section{Related Work}\label{sec:relatedwork}

A common approach to relational verification and program equivalence
is the use of product programs~\cite{productprograms}. Product
programs combine the states of two programs and interleave their
behavior in a single program. \emph{RVT}~\cite{rvt} proves the
equivalence of C programs by combining them in a product program. By
assuming that the program states are equal after each loop iteration,
\emph{RVT} avoids the need for user-specified or inferred loop
invariants and coupling predicates.

Hawblitzel~et~al.~\cite{mutualsummaries} use a
similar technique for handling recursive function calls.
Felsing~et~al.~\cite{automatingregver} demonstrate that coupling
predicates for proving the equivalence of two programs can often be
inferred automatically. While the structure of imperative and
\mapreduce{} algorithms tends to be quite different, splitting the
translation into intermediate steps yields programs which are often
structurally similar. We have found that in this case, techniques such
as coupling predicates arise naturally and are useful for selected
parts of an equivalence proof.

Radoi~et~al.~\cite{translatingimperative} describe an automatic
translation of imperative algorithms to \mapreduce{} algorithms based
on rewrite rules. While the rewrite rules are very similar to the ones
used in our approach, we complement rewrite rules by coupling
predicates.
Furthermore we are able to prove equivalence for algorithms for which
the automatic translation from Radoi~et~al. is not capable of
producing efficient \mapreduce{} algorithms.
The objective of verification imposes different constraints than the
automated translation -- in particular both programs are provided by
the user, so there is less flexibility needed in the formulation of rewrite
rules.

Chen~et~al.~\cite{sparkspecification} and
Radoi~et~al.~\cite{translatingimperative} describe languages and
sequential semantics for \mapreduce{} algorithms. Chen~et~al.\ describe an
executable sequential specification in the Haskell programming
language focusing on capturing non-determinism correctly. Radoi~et~al.\ use
a language based on a lambda calculus as the common representation for
the previously described translation from imperative to \mapreduce{}
algorithms. While this language closely resembles the language used in
our approach, it lacks support for representing some imperative
constructs such as arbitrary \emph{while}-loops.

Grossman~et~al.~\cite{equivalencespark} verify the equivalence of a
restricted subset of Spark programs by reducing the problem of
checking program equivalence to the validity of formulas in a
decidable fragment of first-order logic. While this approach is fully
automatic, it limits programs to Presburger arithmetic and requires
that they are synchronized in some way.

To the best of our knowledge, we are the first to propose a framework
for proving equivalence of \mapreduce{} and imperative
programs.

\section{Conclusion}\label{sec:conclusion}
We have presented a new approach for proving the equivalence of
imperative and \mapreduce{} algorithms. This approach relies on
splitting the transformation into a chain of intermediate programs.
The individual equivalence proofs are then categorized in
context-independent and context-dependent transformations. Equivalence proofs
for context-independent transformations are handled using rewrite
rules, while equivalence proofs for context-dependent transformations
are based on coupling predicates. We have demonstrated the
feasibility of end-to-end equivalence proofs using this approach by
applying it two well-known non-trivial algorithms.

While we have hinted at the potential for automating this approach,
implementing automation is left as future work. In particular,
it would be interesting to explore whether existing tools for relational
verification using coupling predicates can be used or if new tools are
necessary. Further future work includes extending the approach
presented here to support the full expressiveness provided by
languages which are used to implement imperative and \mapreduce{}
algorithms.

\bibliographystyle{splncs03}
\bibliography{references.bib}

\begin{thebibliography}{10}
\providecommand{\url}[1]{\texttt{#1}}
\providecommand{\urlprefix}{URL }

\bibitem{productprograms}
Barthe, G., Crespo, J.M., Kunz, C.: Relational Verification Using Product
  Programs, pp. 200--214. Springer Berlin Heidelberg, Berlin, Heidelberg
  (2011), \url{http://dx.doi.org/10.1007/978-3-642-21437-0_17}

\bibitem{Thrill}
Bingmann, T., Axtmann, M., J{\"{o}}bstl, E., Lamm, S., Nguyen, H.C., Noe, A.,
  Schlag, S., Stumpp, M., Sturm, T., Sanders, P.: {Thrill}: High-performance
  algorithmic distributed batch data processing with {C++}. In: IEEE
  International Conference on Big Data. pp. 172--183. IEEE (Dec 2016),
  \url{https://doi.org/10.1109/BigData.2016.7840603}, preprint arXiv:1608.05634

\bibitem{pagerank}
Brin, S., Page, L.: The anatomy of a large-scale hypertextual web search
  engine. Comput. Netw. ISDN Syst.  30(1-7),  107--117 (Apr 1998),
  \url{http://dx.doi.org/10.1016/S0169-7552(98)00110-X}

\bibitem{chambers2010flumejava}
Chambers, C., Raniwala, A., Perry, F., Adams, S., Henry, R.R., Bradshaw, R.,
  Weizenbaum, N.: {FlumeJava}: Easy, efficient data-parallel pipelines. {ACM}
  {SIGPLAN} Notices  45(6),  363--375 (2010)

\bibitem{sparkspecification}
Chen, Y.F., Hong, C.D., Lengál, O., Mu, S.C., Sinha, N., Wang, B.Y.: An
  executable sequential specification for spark aggregation (2017)

\bibitem{commutativityofreducers}
Chen, Y.F., Hong, C.D., Sinha, N., Wang, B.Y.: Commutativity of Reducers, pp.
  131--146. Springer (2015),
  \url{http://dx.doi.org/10.1007/978-3-662-46681-0_9}

\bibitem{commutativitytransducer}
Chen, Y., Song, L., Wu, Z.: The commutativity problem of the mapreduce
  framework: {A} transducer-based approach. CoRR  abs/1605.01497 (2016),
  \url{http://arxiv.org/abs/1605.01497}

\bibitem{mapreduce}
Dean, J., Ghemawat, S.: {MapReduce}: Simplified data processing on large
  clusters. Commun. ACM  51(1),  107--113 (Jan 2008),
  \url{http://doi.acm.org/10.1145/1327452.1327492}

\bibitem{mutualtermination}
Elenbogen, D., Katz, S., Strichman, O.: Proving mutual termination. Form.
  Methods Syst. Des.  47(2),  204--229 (Oct 2015),
  \url{http://dx.doi.org/10.1007/s10703-015-0234-3}

\bibitem{automatingregver}
Felsing, D., Grebing, S., Klebanov, V., R\"{u}mmer, P., Ulbrich, M.: Automating
  regression verification. In: Proceedings of the 29th ACM/IEEE International
  Conference on Automated Software Engineering. pp. 349--360. ASE '14, ACM, New
  York, NY, USA (2014), \url{http://doi.acm.org/10.1145/2642937.2642987}

\bibitem{rvt}
Godlin, B., Strichman, O.: Regression verification. In: Proceedings of the 46th
  Annual Design Automation Conference. pp. 466--471. DAC '09, ACM, New York,
  NY, USA (2009), \url{http://doi.acm.org/10.1145/1629911.1630034}

\bibitem{equivalencespark}
Grossman, S., Cohen, S., Itzhaky, S., Rinetzky, N., Sagiv, M.: Verifying
  Equivalence of Spark Programs, pp. 282--300. Springer International
  Publishing, Cham (2017), \url{https://doi.org/10.1007/978-3-319-63390-9_15}

\bibitem{mutualsummaries}
Hawblitzel, C., Kawaguchi, M., Lahiri, S., Reb\^elo, H.: Mutual summaries:
  Unifying program comparison techniques. In: Informal proceedings of BOOGIE
  2011 workshop (2011),
  \url{https://www.microsoft.com/en-us/research/publication/mutual-summaries-unifying-program-comparison-techniques/}

\bibitem{bigstep}
Kahn, G.: Natural semantics. STACS 87 pp. 22--39 (1987)

\bibitem{symdiff}
Lahiri, S.K., Hawblitzel, C., Kawaguchi, M., Reb\^{e}lo, H.: Symdiff: A
  language-agnostic semantic diff tool for imperative programs. In: Proceedings
  of the 24th International Conference on Computer Aided Verification. pp.
  712--717. CAV'12, Springer-Verlag, Berlin, Heidelberg (2012),
  \url{http://dx.doi.org/10.1007/978-3-642-31424-7_54}

\bibitem{lloyd1982least}
Lloyd, S.: Least squares quantization in {PCM}. {IEEE} Transactions on
  Information Theory  28(2),  129--137 (1982),
  \url{https://doi.org/10.1109/TIT.1982.1056489}

\bibitem{McCarthy1963}
McCarthy, J.: A basis for a mathematical theory of computation1). In: Braffort,
  P., Hirschberg, D. (eds.) Computer Programming and Formal Systems, Studies in
  Logic and the Foundations of Mathematics, vol.~35, pp. 33 -- 70. Elsevier
  (1963),
  \url{http://www.sciencedirect.com/science/article/pii/S0049237X08720184}

\bibitem{translatingimperative}
Radoi, C., Fink, S.J., Rabbah, R., Sridharan, M.: Translating imperative code
  to mapreduce. SIGPLAN Not.  49(10),  909--927 (Oct 2014),
  \url{http://doi.acm.org/10.1145/2714064.2660228}

\bibitem{Coq}
development team, T.C.: The Coq proof assistant reference manual. LogiCal
  Project (2004), \url{http://coq.inria.fr}, version 8.6

\bibitem{verdoolaege}
Verdoolaege, S., Janssens, G., Bruynooghe, M.: Equivalence checking of static
  affine programs using widening to handle recurrences. ACM Trans. Program.
  Lang. Syst.  34(3),  11:1--11:35 (Nov 2012),
  \url{http://doi.acm.org/10.1145/2362389.2362390}

\bibitem{white2012hadoop}
White, T.: Hadoop: The definitive guide. O'Reilly Media, Inc. (2012)

\bibitem{spark}
Zaharia, M., Chowdhury, M., Franklin, M.J., Shenker, S., Stoica, I.: Spark:
  Cluster computing with working sets. In: Proceedings of the 2Nd USENIX
  Conference on Hot Topics in Cloud Computing. pp. 10--10. HotCloud'10, USENIX
  Association, Berkeley, CA, USA (2010),
  \url{http://dl.acm.org/citation.cfm?id=1863103.1863113}

\end{thebibliography}

\clearpage

\appendix

\section{Syntax and Semantics of \FFL{}}\label{sec:ffl-appendix}

\begin{figure}[b!]
  \[
  \begin{array}{rclllllll}
    b & := & \mathsf{true} \alt \mathsf{false} \\
    i & := & \multicolumn{4}{l}{\ldots \alt {-2} \alt {-1} \alt 0 \alt 1 \alt 2 \alt \ldots} \\
    e & :=   & x                 &      &                  &      &                    &      &                       \\
      & \alt & \tapp{e}{e}       & \alt & \tabs{x}{e}      &      &                    &      &                       \\
      & \alt & \tint{i}          & \alt & \tbool{b}        & \alt & \tlist{e,\ldots,e} &      &                       \\
      & \alt & \tadd{e}{e}       & \alt & \tsub{e}{e}      & \alt & \tmul{e}{e}        &      &                       \\
      & \alt & \tgt{e}{e}        & \alt & \tlt{e}{e}       &      &                    &      &                       \\
      & \alt & \tunit{}          & \alt & \tpair{e}{e}     &      &                    &      &                       \\
      & \alt & \tfst{e}          & \alt & \tsnd{e}         &      &                    &      &                       \\
      & \alt & \tinl{e}          & \alt & \tinr{e}         & \alt & \multicolumn{3}{l}{\tcase{e}{e}{e}}               \\
      & \alt & \titer{e}{e}      & \alt & \tfold{e}{e}{e}  & \alt & \tif{e}{e}{e}      &      &                       \\
      & \alt & \tread{e}{e}      & \alt & \twrite{e}{e}{e} & \alt & \treadatkey{e}{e}  & \alt & \twriteatkey{e}{e}{e} \\
      & \alt & \treplicate{e}{e} & \alt & \trange{e}{e} & \alt & \tlength{e}        &      &                       \\
      & \alt & \tmap{e}{e}       & \alt & \tgroup{e}       & \alt & \tzip{e}{e}        & \alt & \tconcat{e}           \\
\end{array}
  \]
  \caption{Grammar for terms $e$ of \FFL.}\label{fig:langterms}
\end{figure}

\subsection{Syntax}\label{sec:ffl-syntax}

The programming language \FFL{} is used in our approach to represent
the imperative input program, the target \mapreduce{}
implementations, and intermediate programs in which both programming
paradigms appear in combination.
\FFL{} is designed to accommodate both functional and imperative
concepts. It bases on
\begin{enumerate}
\item \emph{simply typed lambda calculus} with the following theories:
\item \emph{direct sums} and \emph{products},
\item (mathematical) \emph{integers},
\item McCarthy's theory of \emph{arrays}~\cite{McCarthy1963},
\item \emph{iterating} primitives to represent imperative loops,
\item and primitives found in \mapreduce{} frameworks.
\end{enumerate}

\reffig{langterms} shows an exhaustive list of the constructors of
\FFL{}. These can be classified according to the six categories mentioned above.
\begin{description}
\item[Simply typed lambda calculus]~
  \begin{itemize}
  \item variables $x$, function abstraction $\tabs{x}{e}$ and
    application $\tapp{e_{1}}{e_{2}}$
  \item $\tif{e_{1}}{e_{t}}{e_{f}}$
  \item Boolean literals $\tbool{b}$ for
    $b \in \{\mathsf{true},\mathsf{false}\}$
  \end{itemize}

\item[Direct sums and products]~
  \begin{itemize}
  \item introduction $\tpair{e_{1}}{e_{2}}$ and elimination
    forms $\tfst{e}$, $\tsnd{e}$ of binary product types
  \item introduction $\tinl{e}$, $\tinr{e}$, and elimination forms
    \[\tcase{e}{{e_{l}}}{{e_{r}}}\] of binary sum types
  \item $\tunit$, the single inhabitant of $\TUnit{}$
  \end{itemize}

\item[Theory of integers]~
  \begin{itemize}
  \item integer literals $\tint{i}$ for $i\in\mathbb{Z}$
  \item comparison of integers $\tgt{e_{1},e_{2}}$, $\tlt{e_{1},e_{2}}$
  \item binary operations on integers $\tadd{e_{1}}{e_{2}}$, $\tsub{e_{1}}{e_{2}}$, $\tmul{e_{1}}{e_{2}}$
  \end{itemize}

\item[Theory of arrays]~
  \begin{itemize}
  \item array literals $\tlist{e_{1},\ldots,e_{n}}$
  \item array read $\tread{e_{1}}{e_{2}}$
  \item array write $\twrite{e_{1}}{e_{2}}{e_{3}}$
  \item $\treplicate{e_{1}}{e_{2}}$ which produces an array of the
    specified length $e_1$ containing the same element $e_2$ at each
    index,
  \item $\trange{e_{1}}{e_{2}}$ which produces an array containing all
    elements in the range from $e_1$ to $e_2$ and
  \item $\tlength{e}$ which returns the length of array $e$.
  \end{itemize}

\item[Iteration functions]~
  \begin{itemize}
  \item $\titer{e_{1}}{e_{2}}$ and
  \item $\tfold{e_{1}}{e_{2}}{e_{3}}$
  \end{itemize}

\item[\mapreduce{} primitives]~
  \begin{itemize}
  \item $\tmap{f}{\mathit{xs}}$ which applies a function $f$ to all
    elements in the array $\mathit{xs}$
  \item $\tzip{\mathit{xs}}{\mathit{ys}}$ which combines two arrays
    $\mathit{xs},\mathit{ys}$ of the same length into an array
    containing pairs of elements of $\mathit{xs}$ and $\mathit{ys}$
  \item $\tconcat{\mathit{xss}}$ which flattens an array of arrays by
    concatenating the arrays

  \item operations which operate on arrays of key-value pairs:
    $\treadatkey{xs}{k}$ returns the value associated with a key
  \item $\twriteatkey{xs}{k}{v}$ sets the value associated with a key,
    and $\tgroup{xs}$ produces an array that associates each key in
    the original array with all values associated with
    it
  \item $\mathsf{readAtKey}$ and $\mathsf{writeAtKey}$ operate on the
    first pair with a matching key; for $\mathsf{group}$ the order of
    the keys in the resulting array matches the order of their first
    occurences in the input and the order of the values associated
    with each key matches their order in the input array
  \end{itemize}
\end{description}

Fig.~\ref{fig:langtyping} lists the typing rules for \FFL{}. The set
of expressions derivable using the given rules is the set of
well-typed programs in \FFL. $\Gamma$ denotes the context (type
assignment for free variables) under which the rules apply. The
following type constructors are used in the figure.
\begin{center}
  \begin{tabular}{ll}
    \toprule
    Type & Terms of this type \\
    \midrule
    $\TInt{}/\TBool{}$ & integers/booleans \\
    $\TProd{\alpha}{\beta}$ & pairs of terms of type $\alpha$ and $\beta$\\
    $\TSum{\alpha}{\beta}$ & elements of type $\alpha$ or $\beta$ (direct sum)\\
    $\TArrow{\alpha}{\beta}$ & functions from terms of type $\alpha$ to terms of type $\beta$ \\
    $\TList{\alpha}$ & arrays of terms of type $\alpha$ \\
    $\TUnit$ & $\tunit$ \\
    \bottomrule
  \end{tabular}
\end{center}

\begin{figure}[p]
  \scalebox{.8}{
  \renewcommand{\arraystretch}{2.5}
  \centering
  \begin{tabular}{@{}ll@{}}
    \bottomAlignProof{}
      \AXC{}
      \UIC{\hastype{\Gamma,x:\tau}{x}{\tau}}
      \DP{}
    & \bottomAlignProof{}
      \AXC{\hastype{\Gamma}{f}{\TArrow{\alpha}{\beta}}}
      \AXC{\hastype{\Gamma}{t}{\alpha}}
      \BIC{\hastype{\Gamma}{\tapp{f}{t}}{\beta}}
      \DP{}
    \\
      \bottomAlignProof{}
      \AXC{\hastype{\Gamma,x:\alpha}{e}{\beta}}
      \UIC{\hastype{\Gamma}{\tabs{x}{e}}{\TArrow{\alpha}{\beta}}}
      \DP{}
    &
      \bottomAlignProof{}
      \AXC{\hastype{\Gamma}{t_{1}}{\tau}}
      \AXC{\(\ldots\)}
      \AXC{\hastype{\Gamma}{t_{n}}{\tau}}
      \TIC{\hastype{\Gamma}{\tlist{t_{1},\ldots,t_{n}}}{\TList{\tau}}}
      \DP{}
    \\
      \bottomAlignProof{}
      \AXC{\(b\in\{\mathsf{true},\mathsf{false}\}\)}
      \UIC{\hastype{\Gamma}{\tbool{b}}{\TBool}}
      \DP{}
    &
      \bottomAlignProof{}
      \AXC{\(i\in\mathbb{Z}\)}
      \UIC{\hastype{\Gamma}{\tint{i}}{\TInt}}
      \DP{}
      \\
      \bottomAlignProof{}
    \AXC{\hastype{\Gamma}{a}{\TInt}}
    \AXC{\hastype{\Gamma}{b}{\TInt}}
    \BIC{\hastype{\Gamma}{\tadd{a}{b}}{\TInt}}
    \DP{}
    &
      \bottomAlignProof{}
      \AXC{\hastype{\Gamma}{a}{\TInt}}
      \AXC{\hastype{\Gamma}{b}{\TInt}}
      \BIC{\hastype{\Gamma}{\tsub{a}{b}}{\TInt}}
      \DP{}
    \\
    \bottomAlignProof{}
    \AXC{\hastype{\Gamma}{a}{\TInt}}
    \AXC{\hastype{\Gamma}{b}{\TInt}}
    \BIC{\hastype{\Gamma}{\tmul{a}{b}}{\TInt}}
    \DP{} \\
    \bottomAlignProof{}
    \AXC{\hastype{\Gamma}{a}{\TInt}}
    \AXC{\hastype{\Gamma}{b}{\TInt}}
    \BIC{\hastype{\Gamma}{\tgt{a}{b}}{\TBool}}
    \DP{}
    &
      \bottomAlignProof{}
      \AXC{\hastype{\Gamma}{a}{\TInt}}
      \AXC{\hastype{\Gamma}{b}{\TInt}}
      \BIC{\hastype{\Gamma}{\tlt{a}{b}}{\TBool}}
      \DP{}
    \\
      \bottomAlignProof{}
      \AXC{\hastype{\Gamma}{b}{\TBool}}
      \AXC{\hastype{\Gamma}{e_{t}}{\tau}}
      \AXC{\hastype{\Gamma}{e_{f}}{\tau}}
      \TIC{\hastype{\Gamma}{\tif{b}{e_{t}}{e_{f}}}{\tau}}
      \DP{}
    &
      \bottomAlignProof{}
      \AXC{}
      \UIC{\hastype{\Gamma}{\tunit}{\TUnit}}
      \DP{}
    \\
      \bottomAlignProof{}
      \AXC{\hastype{\Gamma}{\mathit{ts}}{\TList{\tau}}}
      \UIC{\hastype{\Gamma}{\tlength{\mathit{ts}}}{\TInt}}
      \DP{}
      &
      \bottomAlignProof{}
      \AXC{\hastype{\Gamma}{a}{\alpha}}
      \AXC{\hastype{\Gamma}{b}{\beta}}
      \BIC{\hastype{\Gamma}{\tpair{a}{b}}{\TProd{\alpha}{\beta}}}
      \DP{}
    \\
      \bottomAlignProof{}
      \AXC{\hastype{\Gamma}{p}{\TProd{\alpha}{\beta}}}
      \UIC{\hastype{\Gamma}{\tfst{p}}{\alpha}}
      \DP{}
    &
      \bottomAlignProof{}
      \AXC{\hastype{\Gamma}{p}{\TProd{\alpha}{\beta}}}
      \UIC{\hastype{\Gamma}{\tsnd{p}}{\beta}}
      \DP{}
    \\
      \bottomAlignProof{}
      \AXC{\hastype{\Gamma}{a}{\alpha}}
      \UIC{\hastype{\Gamma}{\tinl{a}}{\TSum{\alpha}{\beta}}}
      \DP{}
    &
      \bottomAlignProof{}
      \AXC{\hastype{\Gamma}{b}{\beta}}
      \UIC{\hastype{\Gamma}{\tinr{b}}{\TSum{\alpha}{\beta}}}
      \DP{}
    \\
    \multicolumn{2}{l}{
      \bottomAlignProof{}
      \AXC{\hastype{\Gamma}{e}{\TSum{\alpha}{\beta}}}
      \AXC{\hastype{\Gamma,l:\alpha}{a}{\tau}}
      \AXC{\hastype{\Gamma,r:\beta}{b}{\tau}}
      \TIC{\hastype{\Gamma}{\tcase{e}{a}{b}}{\tau}}
    \DP{}
    }
    \\
    \multicolumn{2}{l}{
      \bottomAlignProof{}
      \AXC{\hastype{\Gamma}{f}{\TArrow{\TProd{\alpha}{\beta}}{\alpha}}}
      \AXC{\hastype{\Gamma}{e}{\alpha}}
      \AXC{\hastype{\Gamma}{\mathit{ts}}{\TList{\beta}}}
      \TIC{\hastype{\Gamma}{\tfold{f}{e}{\mathit{ts}}}{\alpha}}
    \DP{}
    }
    \\
    \multicolumn{2}{l}{
      \bottomAlignProof{}
      \AXC{\hastype{\Gamma}{f}{\TArrow{\alpha}{\TSum{\TUnit}{\alpha}}}}
      \AXC{\hastype{\Gamma}{e}{\alpha}}
      \BIC{\hastype{\Gamma}{\titer{f}{e}}{\alpha}}
    \DP{}
    }
    \\
      \bottomAlignProof{}
      \AXC{\hastype{\Gamma}{\mathit{ts}}{\TList{\tau}}}
      \AXC{\hastype{\Gamma}{i}{\TInt}}
      \BIC{\hastype{\Gamma}{\tread{\mathit{ts}}{i}}{\tau}}
      \DP{}
    &
      \bottomAlignProof{}
      \AXC{\hastype{\Gamma}{\mathit{ts}}{\TList{\tau}}}
      \AXC{\hastype{\Gamma}{i}{\TInt}}
      \AXC{\hastype{\Gamma}{v}{\tau}}
      \TIC{\hastype{\Gamma}{\twrite{\mathit{ts}}{i}{v}}{\TList{\tau}}}
      \DP{}
    \\
      \bottomAlignProof{}
      \AXC{\hastype{\Gamma}{\mathit{ts}}{\TList{\TProd{\alpha}{\beta}}}}
      \AXC{\hastype{\Gamma}{k}{\alpha}}
      \BIC{\hastype{\Gamma}{\treadatkey{\mathit{ts}}{k}}{\TSum{\tunit}{\beta}}}
      \DP{}
    &
      \bottomAlignProof{}
      \AXC{\hastype{\Gamma}{\mathit{ts}}{\TList{\TProd{\alpha}{\beta}}}}
      \AXC{\hastype{\Gamma}{k}{\alpha}}
      \AXC{\hastype{\Gamma}{t}{\beta}}
      \TIC{\hastype{\Gamma}{\twriteatkey{\mathit{ts}}{k}{t}}{\TList{\TProd{\alpha}{\beta}}}}
      \DP{}
    \\
      \bottomAlignProof{}
      \AXC{\hastype{\Gamma}{n}{\TInt}}
      \AXC{\hastype{\Gamma}{t}{\alpha}}
      \BIC{\hastype{\Gamma}{\treplicate{n}{t}}{\TList{\alpha}}}
      \DP{}
    &
      \bottomAlignProof{}
      \AXC{\hastype{\Gamma}{l}{\TInt}}
      \AXC{\hastype{\Gamma}{r}{\TInt}}
      \AXC{\hastype{\Gamma}{s}{\TInt}}
      \TIC{\hastype{\Gamma}{\trange{l}{r}}{\TList{\TInt}}}
      \DP{}
    \\
    &
      \bottomAlignProof{}
      \AXC{\hastype{\Gamma}{f}{\TArrow{\alpha}{\beta}}}
      \AXC{\hastype{\Gamma}{\mathit{ts}}{\TList{\alpha}}}
      \BIC{\hastype{\Gamma}{\tmap{f}{\mathit{ts}}}{\TList{\beta}}}
      \DP{}
    \\
      \bottomAlignProof{}
      \AXC{\hastype{\Gamma}{\mathit{ts}}{\TList{\TProd{\alpha}{\beta}}}}
      \UIC{\hastype{\Gamma}{\tgroup{\mathit{ts}}}{\TList{\TProd{\alpha}{\TList{\beta}}}}}
      \DP{}
    &
      \bottomAlignProof{}
      \AXC{\hastype{\Gamma}{\mathit{as}}{\TList{\alpha}}}
      \AXC{\hastype{\Gamma}{\mathit{bs}}{\TList{\beta}}}
      \BIC{\hastype{\Gamma}{\tzip{\mathit{as}}{\mathit{bs}}}{\TList{\TProd{\alpha}{\beta}}}}
      \DP{}
    \\
      \bottomAlignProof{}
      \AXC{\hastype{\Gamma}{\mathit{tss}}{\TList{\TList{\alpha}}}}
      \UIC{\hastype{\Gamma}{\tconcat{\mathit{tss}}}{\TList{\alpha}}}
      \DP{}
  \end{tabular}}
  \caption{Typing rules for \FFL.}\label{fig:langtyping}
\end{figure}

\subsection{Semantics}
\label{sec:ffl-semantics}

The semantics of \FFL{} is given as a reduction big-step semantics
formalized as a binary predicate $\steprel$ on well-typed and closed
\FFL\ terms. The values (terms which evaluate to themselves) in \FFL{}
are: $\lambda$-abstractions, literals, and sums and products of
values. Fig.~\ref{fig:langsemantics} lists the evaluation rules for
most constructs and Fig.~\ref{fig:semantics-keyvalue} shows those for
the key-value operations in \FFL.

\begin{figure}[b!]
  \scalebox{.8}{
  \renewcommand{\arraystretch}{3}
  \begin{tabular}{@{}c@{}}
    \bottomAlignProof{}
    \AXC{\steps{\mathit{xs}}{[(k_{1},v_{1}),\ldots,(k_{j},v_{j}),\ldots,(k_{n},v_{n})]}}
    \AXC{$\forall 1\leq i<j, k_{i}\neq k$}
    \AXC{$k_{j}=k$}
    \TIC{\steps{\treadatkey{\mathit{xs}}{k}}{\tinr{v}}}
    \DP{}
    \\
    \bottomAlignProof{}
    \AXC{\steps{\mathit{xs}}{[(k_{1},v_{1}),\ldots,(k_{n},v_{n})]}}
    \AXC{$\forall 1\leq i\leq n, k_{i}\neq k$}
    \BIC{\steps{\treadatkey{\mathit{xs}}{k}}{\tinl{\tunit}}}
    \DP{}
    \\
    \bottomAlignProof{}
    \AXC{\steps{\mathit{xs}}{[(k_{1},v_{1}),\ldots,(k_{j},v_{j}),\ldots,(k_{n},v_{n})]}}
    \AXC{$\forall 1\leq i<j, k_{i}\neq k$}
    \AXC{$k_{j}=k$}
    \TIC{\steps{\twriteatkey{\mathit{xs}}{k}{v}}{[(k_{1},v_{1}),\ldots,(k_{j},v),\ldots,(k_{n},v_{n})]}}
    \DP{}
    \\
    \bottomAlignProof{}
    \AXC{\steps{\mathit{xs}}{[(k_{1},v_{1}),\ldots,(k_{n},v_{n})]}}
    \AXC{$\forall 1\leq i \leq n, k_{i}\neq k$}
    \BIC{\steps{\twriteatkey{\mathit{xs}}{k}{v}}{[(k_{1},v_{1}),\ldots,(k_{n},v_{n}),(k,v)]}}
    \DP{}
    \\
    \bottomAlignProof{}
    \AXC{\steps{\mathit{xs}}{[]}}
    \UIC{\steps{\tgroup{\mathit{xs}}}{[]}}
    \DP{}
    \\
    \bottomAlignProof{}
    \AXC{\steps{\mathit{xs}}{[(k_{1},v_{1}),\ldots,(k_{n},v_{n})]}}
    \AXC{
      \stackanchor{\(\tgroup{[(k_{1},v_{1}),\ldots,(k_{n-1},v_{n-1})]} \steprel\)}
                  {\([(k'_{1},vs_{1}),\ldots,(k'_{m},vs_{m})]\)}}
    \AXC{$\forall 1\leq i\leq m, k_{n}\neq k'_{i}$}
    \TIC{\steps{\tgroup{\mathit{xs}}}{[(k'_{1},vs_{1}),\ldots,(k'_{m},vs_{m}),(k_{n},[v_{n}])]}}
    \DP{}
    \\
    \def\defaultHypSeparation{\hskip.1cm}
.
    \bottomAlignProof{}
    \AXC{\stackanchor{\(\mathit{xs}\steprel\)}{\([(k_{1},v_{1}),\ldots,(k_{n},v_{n})]\)}}
    \AXC{
    \stackanchor%
    {\(\tgroup{[(k_{1},v_{1}),\ldots,(k_{n},v_{n})]}\steprel\)}
    {\([(k'_{1},vs_{1}),\ldots,(k_{j},[v'_{1},\ldots,v'_{n}]),\ldots,(k'_{n},vs_{n})]\)}}
    \AXC{\(\forall 1\leq i < j, k'_{n}\neq k_{i}\)}
    \AXC{\(k_{j}=k\)}
    \QIC{\steps{\tgroup{\mathit{xs}}}{[(k'_{1},vs_{1}),\ldots,(k,[v'_{1},\ldots,v'_{n},v_{n}]),\ldots,(k_{n},[v_{n}])]}}
    \DP{}
  \end{tabular}}
\caption{Semantics of key-value operations in \FFL{} (for
  readability, we use the shorthands
  \([t_{1},\ldots,t_{n}]\) for list literals
  \(\tlist{t_{1},\ldots,t_{n}}\), and \((t_{1},t_{2})\) for pairs
  \(\tpair{t_{1}}{t_{2}}\)).}\label{fig:semantics-keyvalue}
\end{figure}

\begin{figure}
  \scalebox{.8}{
  \renewcommand{\arraystretch}{3}
  \centering
  \begin{tabular}{ll}
      \bottomAlignProof{}
      \AXC{\steps{f}{\tabs{x}{b}}}
      \AXC{\steps{t}{t'}}
      \AXC{\steps{\subst{t'}{x}{b}}{v}}
      \TIC{\steps{\tapp{f}{t}}{v}}
      \DP{}
    &
      \bottomAlignProof{}
      \AXC{}
      \UIC{\steps{\tabs{x}{e}}{\tabs{x}{e}}}
      \DP{}
    \\
      \bottomAlignProof{}
      \AXC{}
      \UIC{\steps{\tbool{b}}{\tbool{b}}}
      \DP{}
    &
      \bottomAlignProof{}
      \AXC{}
      \UIC{\steps{\tint{i}}{\tint{i}}}
      \DP{}
    \\
      \bottomAlignProof{}
      \AXC{\steps{t_{1}}{t_{1}'}}
      \AXC{\(\ldots\)}
      \AXC{\steps{t_{n}}{t_{n}'}}
      \TIC{\steps{\tlist{t_{1},\ldots,t_{n}}}{\tlist{t_{1}',\ldots,t_{n}'}}}
      \DP{}
    &
      \bottomAlignProof{}
      \AXC{\steps{\mathit{ts}}{\tlist{t_{1},\ldots,t_{n}}}}
      \UIC{\steps{\tlength{\mathit{ts}}}{\tint{n}}}
      \DP{}
    \\
      \bottomAlignProof{}
      \AXC{\steps{b}{\tbool{\mathsf{true}}}}
      \AXC{\steps{e_{t}}{e_{t'}}}
      \BIC{\steps{\tif{b}{e_{t}}{e_{f}}}{e_{t'}}}
      \DP{}
    &
      \bottomAlignProof{}
      \AXC{\steps{b}{\tbool{\mathsf{false}}}}
      \AXC{\steps{e_{f}}{e_{f'}}}
      \BIC{\steps{\tif{b}{e_{t}}{e_{f}}}{e_{f'}}}
      \DP{}
    \\
    \bottomAlignProof{}
    \AXC{\steps{a}{\tint{i}}}
    \AXC{\steps{b}{\tint{j}}}
    \BIC{\steps{\tadd{a}{b}}{\tint{i+j}}}
    \DP{}
    &
      \bottomAlignProof{}
      \AXC{\steps{a}{\tint{i}}}
      \AXC{\steps{b}{\tint{j}}}
      \BIC{\steps{\tsub{a}{b}}{\tint{i-j}}}
      \DP{}
    \\
    \bottomAlignProof{}
    \AXC{\steps{a}{\tint{i}}}
    \AXC{\steps{b}{\tint{j}}}
    \BIC{\steps{\tmul{a}{b}}{\tint{i\cdot j}}}
    \DP{}
    \\
    \bottomAlignProof{}
    \AXC{\steps{a}{\tint{i}}}
    \AXC{\steps{b}{\tint{j}}}
    \BIC{\steps{\tgt{a}{b}}{\tbool{i>j}}}
    \DP{}
    &
      \bottomAlignProof{}
      \AXC{\steps{a}{\tint{i}}}
      \AXC{\steps{b}{\tint{j}}}
      \BIC{\steps{\tlt{a}{b}}{\tbool{i<j}}}
      \DP{}
    \\
      \bottomAlignProof{}
      \AXC{}
      \UIC{\steps{\tunit}{\tunit}}
      \DP{}
    &
      \bottomAlignProof{}
      \AXC{\steps{a}{a'}}
      \AXC{\steps{b}{b'}}
      \BIC{\steps{\tpair{a}{b}}{\tpair{a'}{b'}}}
      \DP{}
    \\
      \bottomAlignProof{}
      \AXC{\steps{p}{\tpair{a}{b}}}
      \UIC{\steps{\tfst{p}}{a}}
      \DP{}
    &
      \bottomAlignProof{}
      \AXC{\steps{p}{\tpair{a}{b}}}
      \UIC{\steps{\tfst{p}}{b}}
      \DP{}
    \\
      \bottomAlignProof{}
      \AXC{\steps{a}{a'}}
      \UIC{\steps{\tinl{a}}{\tinl{a'}}}
      \DP{}
    &
      \bottomAlignProof{}
      \AXC{\steps{b}{b'}}
      \UIC{\steps{\tinr{b}}{\tinr{b'}}}
      \DP{}
    \\
      \bottomAlignProof{}
      \AXC{\steps{e}{\tinl{v}}}
      \AXC{\steps{\subst{v}{l}{a}}{a'}}
      \BIC{\steps{\tcase{e}{a}{b}}{a'}}
      \DP{}
    &
      \bottomAlignProof{}
      \AXC{\steps{e}{\tinr{v}}}
      \AXC{\steps{\subst{v}{r}{b}}{b'}}
      \BIC{\steps{\tcase{e}{a}{b}}{b'}}
      \DP{}
    \\
      \bottomAlignProof{}
      \AXC{\steps{i}{\tint{i'}}}
      \AXC{\steps{\mathit{ts}}{\tlist{t_{1},\ldots,t_{i'},\ldots,t_{n}}}}
      \BIC{\steps{\tread{\mathit{ts}}{i}}{t_{i'}}}
      \DP{}
    \\
      \multicolumn{2}{l}{
      \bottomAlignProof{}
      \AXC{\steps{i}{\tint{i'}}}
      \AXC{\steps{\mathit{ts}}{\tlist{t_{1},\ldots,t_{i'},\ldots,t_{n}}}}
      \AXC{\steps{t}{t'}}
      \TIC{\steps{\twrite{\mathit{ts}}{i}{t}}{\tlist{t_{1},\ldots,t',\ldots,t_{n}}}}
    \DP{}
    }
    \\
      \bottomAlignProof{}
      \AXC{\steps{acc}{acc'}}
      \AXC{\steps{\tapp{f}{acc'}}{\tinl{\tunit}}}
      \BIC{\steps{\titer{f}{acc}}{acc'}}
      \DP{}
    &
      \bottomAlignProof{}
      \AXC{\steps{\tapp{f}{acc}}{\tinr{acc'}}}
      \AXC{\steps{\titer{f}{acc'}}{v}}
      \BIC{\steps{\titer{f}{acc}}{v}}
      \DP{}
    \\
    \multicolumn{2}{l}{
      \bottomAlignProof{}
      \AXC{\steps{f}{\tabs{f}{b}}}
      \AXC{\steps{acc}{acc'}}
      \AXC{\steps{\mathit{ts}}{\tlist{}}}
      \TIC{\steps{\tfold{f}{acc}{\mathit{ts}}}{acc'}}
      \DP{}
    }
    \\
    \multicolumn{2}{l}{
      \bottomAlignProof{}
      \AXC{\steps{\mathit{ts}}{\tlist{t_{1},t_{2},\ldots,t_{n}}}}
      \AXC{\steps{\tapp{f}{\tpair{acc}{t_{1}}}}{acc'}}
      \AXC{\steps{\tfold{f}{acc'}{\tlist{\mathit{t_{2},\ldots,t_{n}}}}}{v}}
      \TIC{\steps{\tfold{f}{acc}{\mathit{ts}}}{v}}
      \DP{}}
    \\
      \bottomAlignProof{}
      \AXC{\steps{n}{\tint{i}}}
      \AXC{\steps{t}{v}}
      \BIC{\steps{\treplicate{n}{v}}{\tlist{\underbrace{v,\ldots,v}_i}}}
      \DP{}
    \\
      \bottomAlignProof{}
      \AXC{\steps{l}{\tint{l'}}}
      \AXC{\steps{h}{\tint{h'}}}
      \BIC{\steps{\trange{l}{h}}{\tlist{\tint{l'},\tint{l'+1},\ldots,\tint{h'-1}}}}
      \DP{}
    \\
      \multicolumn{2}{l}{
      \bottomAlignProof{}
      \AXC{\steps{\mathit{ts}}{\tlist{t_{1},\ldots,t_{n}}}}
      \AXC{\steps{\tapp{f}{t_{1}}}{t_{1}'}}
      \AXC{\ldots}
      \AXC{\steps{\tapp{f}{t_{n}}}{t_{n}'}}
      \QIC{\steps{\tmap{f}{\mathit{ts}}}{\tlist{t_{1}',\ldots,t_{n}'}}}
      \DP{}
      }
    \\
        \bottomAlignProof{}
        \AXC{\steps{\mathit{as}}{\tlist{a_{1},\ldots,a_{n}}}}
        \AXC{\steps{\mathit{bs}}{\tlist{b_{1},\ldots,b_{n}}}}
        \BIC{\steps{\tzip{\mathit{as}}{\mathit{bs}}}{\tlist{\tpair{a_{1}}{b_{1}},\ldots,\tpair{a_{n}}{b_{n}}}}}
        \DP{}
    \\
      \bottomAlignProof{}
      \AXC{\steps{\mathit{tss}}{\tlist{\tlist{t_{1,1},\ldots,t_{1,n}},\ldots,\tlist{t_{m,1},\ldots,t_{m,n'}}}}}
      \UIC{\steps{\tconcat{\mathit{tss}}}{\tlist{t_{1,1},\ldots,t_{1,n},\ldots,t_{m,1},\ldots,t_{m,n'}}}}
    \DP{}
  \end{tabular}}
  \caption{Big-step semantics of \FFL.}\label{fig:langsemantics}
\end{figure}

\clearpage

\section{Rewrite Rules}\label{sec:rules-appendix}

Below we list all rewrite rules that we have identified. Rule 1 has
been explained previously and deals with extracting parts of the loop
body that are independent of other iterations into a \texttt{map}
expression which allows computing them in parallel. Rules 2 and~3 group
accesses to the same index of an array or the same key in an array of
key-value pairs. Accesses to different indices/keys can then be
computed in parallel. Rule~4 fuses two consecutive applications of
\texttt{map} into a single application. Rule~5 deals with loops that
update an array that they also read from by writing to a separate
array instead. Rule~6 flattens nested \texttt{fold} expressions over
an array of arrays into a single \texttt{fold} expression over the
concatenated arrays. Rules 7 and~8 transform specific kinds of
\texttt{iter} to \texttt{fold} and \texttt{fold} to
\texttt{map}. Rules 9 and~10 transform \texttt{fold} and \texttt{map}
expressions over an index range to \texttt{fold} and \texttt{map}
expressions that operate directly on the values stored at these
indices. Rule~11 commutes writing back updates to an array with
applying a function to all result values. Finally, Rules 12 and~13
commute array reads with \texttt{zip} and \texttt{map}.

\begin{enumerate}[style=nextline,itemsep=5pt, label=\bfseries{\arabic*}.]
\newcommand\litem[2]{\item \textbf{#1.}\par \vspace{1pt}#2}
\litem{Extract independent part of loop body to \texttt{map}}
\rewriterule%
{ \arraycolsep=0pt
  \(
  \begin{array}{ll}
    \mathsf{fold}( & \tabs{\mathit{acc}}{\tabs{x}{f(\mathit{acc},g(x))}}, \\
                   & acc, \\
                   & \mathit{xs})
  \end{array}
  \)
}
{ \arraycolsep=0pt
  \(
  \begin{array}{ll}
    \mathsf{fold}( & \tabs{\mathit{acc}}{\tabs{y}{f(\mathit{acc}, y)}}, \\
                   & acc, \\
                   & \tmap{g}{\mathit{xs}})
  \end{array}
  \)
}
{\({ \mathit{acc}\not\in\fv{f}, \mathit{x}\not\in\fv{f}, \mathit{y}\not\in\fv{f}},\) \\
  \({ \mathit{x}\not\in\fv{g},\mathit{acc}\not\in\fv{g}},g \text{ is not stuck}\)}

\litem{Group accesses to the same index of an array}
\rewriterule%
{ \arraycolsep=0pt
  \(
  \begin{array}{ll}
    \mathsf{fold}( & \lambda\mathit{acc}.\,\lambda(i,x). \\
                   & \twrite{\mathit{acc}}{i}{f(i,x,\mathit{acc}[i])}, \\
                   & ys, \\
                   & \mathit{xs})
  \end{array}
  \)
}
{ \arraycolsep=0pt
  \(
  \begin{array}{ll}
    \mathsf{fold}( &\lambda\mathit{acc}.\,\lambda(i,v).\, \twrite{\mathit{acc}}{i}{v}, \\
                   & \mathit{ys}, \\
                   & \begin{array}{ll}
                       \mathsf{map}( & \lambda(i,\mathit{vs}). \\
                                     & (i,\mathsf{fold}(\begin{array}[t]{l}
                                                          \tabs{x'}{\tabs{x}{f(i,x,x')}}, \\
                                                          \tread{ys}{i}, \\
                                                          \mathit{vs})),
                                                        \end{array} \\
                                     & \tgroup{xs}))
                     \end{array}
  \end{array}
  \)
}
{\\\({\mathit{acc}\not\in\fv{f},\mathit{x}\not\in\fv{f},\mathit{x'}\not\in\fv{f},\mathit{i}\not\in\fv{f},\mathit{vs}\not\in\fv{f}}\)}

\litem{Group accesses to the same key}
\rewriterule%
{ \arraycolsep=0pt
  \(
  \begin{array}{ll}
    \mathsf{fold}( & \lambda\mathit{acc}.\,\lambda(k,v). \\
                   & \mathsf{writeAtKey}( \begin{array}[t]{l}
                                            acc, \\
                                            k, \\
                                            f( \begin{array}[t]{l}
                                                 k, \\ v, \\ \treadatkey{acc}{k})),
                                               \end{array}
                                          \end{array} \\
                   & m, \\
                   & \mathit{xs})
  \end{array}
  \)
}
{ \arraycolsep=0pt
  \(
  \begin{array}{ll}
    \mathsf{fold}( & \begin{array}[t]{l}
                       \lambda\mathit{acc}.\,\lambda(k,v).\, \\
                       \ \twriteatkey{\mathit{acc}}{k}{v},
                     \end{array} \\
                   & m, \\
                   & \begin{array}{ll}
                       \mathsf{map}( & \lambda(k,\mathit{vs}). \\
                                     & (k,\mathsf{fold}(\begin{array}[t]{l}
                                                          \tabs{v'}{\tabs{v}{f(k,v,v')}}, \\
                                                          \treadatkey{m}{k}, \\
                                                          \mathit{vs})),
                                                        \end{array} \\
                                     & \tgroup{xs}))
                     \end{array}
  \end{array}
  \)
}
{\\${\mathit{acc}\not\in\fv{f},\mathit{v}\not\in\fv{f},\mathit{v'}\not\in\fv{f},\mathit{k}\not\in\fv{f},\mathit{vs}\not\in\fv{f}}$}

\litem{Fuse consecutive calls to \texttt{map} into a single call of \texttt{map}}
\rewriterule%
{\(\begin{array}{l}\tmap{f}{\tmap{g}{\mathit{xs}}}\end{array}\)}
{\(\begin{array}{l}\tmap{\tabs{x}{\tapp{f}{\tapp{g}{x}}}}{\mathit{xs}}\end{array}\)}
{\({x\not\in\fv{f},x\not\in\fv{g},f\text{ is not stuck},g\text{ is not stuck}}\)}

\litem{Separate arrays that are read from and written to}
\rewriterule%
{ \arraycolsep=0pt
  \(
  \begin{array}{ll}
    \mathsf{fold}( & \lambda(xs', i).\, \\
                   & \ \begin{array}[t]{ll}
                       \mathsf{write}( & xs', \\
                                       & i, \\
                                       & \begin{array}{ll}
                                           \mathsf{app}( & f, \\
                                                         & (i, \tread{xs'}{i} )),
                                         \end{array}
                     \end{array} \\
                   & xs, \\
                   & \trange{0}{\tlength{xs}})
  \end{array}
  \)
}
{ \arraycolsep=0pt
  \(
  \begin{array}{ll}
    \mathsf{fold}( & \lambda(ys', i).\, \\
                   & \ \begin{array}[t]{ll}
                         \mathsf{write}( & ys', \\
                                         & i, \\
                                         & \begin{array}{ll}
                                             \mathsf{app}( & f, \\
                                                           & (i,\tread{xs}{i})),
                                           \end{array}
                       \end{array} \\
                   & ys \\
                   & \trange{0}{\tlength{xs}})
  \end{array}
  \)
}
{\\\({\tlength{xs}=\tlength{ys},xs'\not\in\fv{f},i\not\in\fv{f},ys'\not\in\fv{xs},i\not\in\fv{xs}}\)}

\litem{Flatten \texttt{fold} over array of arrays}
\rewriterule%
{ \arraycolsep=0pt
  \(
  \begin{array}{ll}
    \tfold{\tabs{(acc', xs)}{\tfold{f}{acc'}{xs}}}{acc}{xss}
  \end{array}
  \)
}
{ \arraycolsep=0pt
  \(
  \begin{array}{ll}
    \tfold{f}{acc}{\tconcat{xss}}
  \end{array}
  \)
}
{\({f\text{ is not stuck}, acc\not\in\fv{f}, xs\not\in\fv{f}}\)}

\litem{Transform \texttt{iter} to \texttt{fold}}
\rewriterule%
{ \arraycolsep=0pt
  \(
  \mathsf{fst}(\mathsf{iter}(\begin{array}[t]{l}
                               \lambda (acc,i).\, \mathsf{if}\: \begin{array}[t]{l}
                                                                i < max \\
                                                                  \mathsf{then}\: \mathsf{inr}(
                                                                  \begin{array}[t]{l}
                                                                    \tapp{f}{(acc, i)}, \\
                                                                    i+1)
                                                                  \end{array} \\
                                                                \mathsf{else}\: \tinl{\tunit},
                                                              \end{array} \\
                               (acc_{0}, min)))
                             \end{array}
  \)
}
{ \arraycolsep=0pt
  \(
  \mathsf{fold}(\begin{array}[t]{l}
                  f, \\
                  acc_{0}, \\
                  \trange{min}{max})
                \end{array}
  \)
}
{ \\ \({f\text{ is not stuck}, acc\not\in\fv{f},i\not\in\fv{f},i\not\in\fv{max},acc\not\in\fv{max}}\) }

\litem{Transform \texttt{fold} to \texttt{map}}
\rewriterule%
{ \arraycolsep=0pt
  \(
  \begin{array}{ll}
    \mathsf{fold}( & \lambda(ys',i).\, \mathsf{twrite} (\begin{array}[t]{l}
                                                          ys', \\
                                                          i, \\
                                                          \tapp{f}{\tread{xs}{i}}),
                                                        \end{array} \\
                   & ys, \\
                   & \trange{0}{\tlength{xs}})
  \end{array}
  \)
}
{\(\tmap{f}{xs}\)}
{\\\({ys'\not\in\fv{f},i\not\in\fv{f},ys'\not\in\fv{xs},i\not\in\fv{xs},f\text{ is not stuck}}\)}

\litem{\texttt{fold} over the values in an array instead of over the index range}
  \rewriterule%
  { \arraycolsep=0pt
    \(
    \begin{array}{ll}
      \mathsf{fold}(\lambda(acc,i).\, & \tapp{f}{(acc,\tread{xs}{i})}, \\
                                      & i, \\
                                      & \trange{i}{\tlength{xs}})
    \end{array}
    \)
  }
  { \(\tfold{f}{i}{xs}\) }
  {\\\({i\not\in\fv{xs},acc\not\in\fv{xs},i\not\in\fv{f},acc\not\in\fv{f}, f\text{ is not stuck}}\)}

\litem{\texttt{map} over the values in an array instead of over the index range}
  \rewriterule%
  { \arraycolsep=0pt
    \(
    \begin{array}{ll}
      \mathsf{map}(\lambda i.\, & \tapp{f}{\tread{xs}{i}}, \\
                                 & \trange{i}{\tlength{xs}})
    \end{array}
    \)
  }
  { \(\tmap{f}{xs}\) }
  {\({i\not\in\fv{xs},i\not\in\fv{f},f\text{ is not stuck}}\)}

\litem{Commute writing back updates to an array and applying \texttt{map} to the result}
\rewriterule%
{ \arraycolsep=0pt
  \(
  \begin{array}{ll}
    \mathsf{map}( & f, \\
                  & \begin{array}{ll}
                      \mathsf{fold}( & \tabs{(xs',(i,x))}{\twrite{xs'}{i}{x}}, \\
                                     & xs, \\
                                     & ys))
                    \end{array} \\
  \end{array}
  \)
}
{ \arraycolsep=0pt
  \(
  \begin{array}{ll}
    \mathsf{fold}( & \tabs{(xs',(i,x))}{\twrite{xs'}{i}{x}}, \\
                   & \tmap{f}{xs}, \\
                   & \begin{array}[t]{ll}
                       \mathsf{map}( & \tabs{(i,x)}{(i,\tapp{f}{x})}, \\
                                     & ys))
                     \end{array}
  \end{array}
  \)
}
{\({i\not\in\fv{f},x\not\in\fv{f}}\)}

\litem{Commute read and zip}
\begin{enumerate}[label={\alph*)}]
  \item
\rewriterule
{
  \( \tfst{\tread{\tzip{xs}{ys}}{i}}
  \)
}
{
  \( \tread{xs}{i}
  \)
}
{ \(\tlength{xs}=\tlength{ys},ys\text{ is not stuck}\)
}
\item
\rewriterule
{
  \( \tsnd{\tread{\tzip{xs}{ys}}{i}}
  \)
}
{
  \( \tread{ys}{i}
  \)
}
{ \(\tlength{xs}=\tlength{ys},xs\text{ is not stuck}\)
}
\end{enumerate}

\litem{Commute read and map}%
\begin{enumerate}[label={\alph*)}]
  \item
\rewriterule%
{
  \( \tread{\tmap{f}{xs}}{i} \)
}
{ \( \tapp{f}{\tread{xs}{i}} \)
}
{}

\item
\rewriterule%
{ \arraycolsep=0pt
  \( \mathsf{readAtKey}(\begin{array}[t]{l}
                          \mathsf{map}(\begin{array}[t]{l}
                                         \lambda (k,v).\\
                                         \ (k, \tapp{f}{v}), \\
                                         xs),
                                         \end{array} \\
                          k)
                        \end{array} \)
}
{ \( \tapp{f}{\treadatkey{xs}{k}} \)
}
{}
\end{enumerate}

\end{enumerate}

\end{document}